\documentclass[12pt,oneside]{amsart}
\usepackage[margin=1in]{geometry}
\usepackage{abstract}
\usepackage{enumitem,verbatim}
\usepackage{multirow}
\usepackage{microtype}
\usepackage[all,cmtip]{xy}
\usepackage{physics}
\usepackage{hyperref}
\usepackage{color}
\hypersetup{colorlinks,linkcolor=blue,citecolor=cyan}
\usepackage{amsthm}
\usepackage[nameinlink]{cleveref}
\usepackage{amsmath,amssymb,stackrel,nicefrac}
\usepackage{amsaddr}
\usepackage{tikz}
\usepackage{pgfplots}
\usepackage{mathtools}

\pgfplotsset{compat=1.10}
\usepgfplotslibrary{fillbetween}
\usetikzlibrary{patterns, arrows, arrows.meta, , decorations.markings}

\newtheorem{prop}{Proposition}
\newtheorem{conj}{Conjecture}
\newtheorem{qn}{Question}
\definecolor{purple}{rgb}{0.7,0,0.7}

\usepackage{graphicx}
\newcommand{\overcross}{
 {\mathchoice
  {\includegraphics[height=1.6ex]{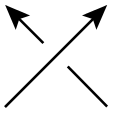}}
  {\includegraphics[height=1.6ex]{overcrossing.png}}
  {\includegraphics[height=1.2ex]{overcrossing.png}}
  {\includegraphics[height=0.9ex]{overcrossing.png}}
 }
}
\newcommand{\undercross}{
 {\mathchoice
  {\includegraphics[height=1.6ex]{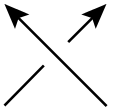}}
  {\includegraphics[height=1.6ex]{undercrossing.png}}
  {\includegraphics[height=1.2ex]{undercrossing.png}}
  {\includegraphics[height=0.9ex]{undercrossing.png}}
 }
}
\newcommand{\nocross}{
 {\mathchoice
  {\includegraphics[height=1.6ex]{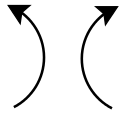}}
  {\includegraphics[height=1.6ex]{no_crossing.png}}
  {\includegraphics[height=1.2ex]{no_crossing.png}}
  {\includegraphics[height=0.9ex]{no_crossing.png}}
 }
}

\newcommand{\R}{{\mathbb{R}}}
\newcommand{\C}{{\mathbb{C}}}
\newcommand{\ind}{\operatorname{index}}

\title{$\widehat{Z}$ at large $N$: from curve counts to quantum modularity}
\author{Tobias Ekholm$^{1,2}$, Angus Gruen$^3$, Sergei Gukov$^3$, Piotr Kucharski$^{3,4}$, Sunghyuk Park$^3$, and Piotr Su{\l}kowski$^{3,4}$}

\address{$^1$Department of Mathematics, Uppsala University, \\ L{\"a}gerhyddsv{\"a}gen 1, 752 37 Uppsala, Sweden}
\address{$^2$Institut Mittag-Leffler, Aurav{\"a}gen 17, 182 60 Djursholm, Sweden}
\address{$^3$Division of Physics, Mathematics and Astronomy, California Institute of Technology, 1200~E.~California Blvd., Pasadena, CA 91125, USA}
\address{$^4$Faculty of Physics, University of Warsaw, ul. Pasteura 5, 02-093 Warsaw, Poland}

\email{tobias.ekholm@math.uu.se}
\email{agruen@caltech.edu}
\email{piotrek@caltech.edu}
\email{spark3@caltech.edu}
\email{psulkows@fuw.edu.pl}

\begin{document}
%\begin{nouppercase}
\maketitle
%\end{nouppercase}
\begin{abstract}
Reducing a~6d fivebrane theory on a~3-manifold $Y$ gives a~$q$-series 3-manifold invariant  $\widehat{Z}(Y)$. %\cite{GPV,GPPV}.
%introduced by Gukov-Pei-Putrov-Vafa. 
We analyse the~large-$N$ behaviour of $F_K=\widehat{Z}(M_K)$, where $M_K$ is the~complement of a~knot $K$ in the~3-sphere,
%(a.k.a.\ $F_{K}$ invariants)
%, which opens the~door for studying quantum modularity in a~new realm. 
and explore the~relationship between an~$a$-deformed ($a=q^N$) version of $F_{K}$ and HOMFLY-PT polynomials. On the~one hand, in combination with counts of holomorphic annuli on knot complements, this gives an~enumerative interpretation of $F_K$ in terms of counts of open holomorphic curves. On the~other, it leads to closed form expressions for $a$-deformed $F_K$ for $(2,2p+1)$-torus knots. They suggest a~further $t$-deformation based on superpolynomials, which can be used to obtain a~$t$-deformation of ADO polynomials, expected to be related to categorification. Moreover, studying how $F_K$ transforms under natural geometric operations on $K$ indicates relations to quantum modularity in a~new setting.
\end{abstract}
\newpage
\tableofcontents

\section{Introduction}

The motivation for the~present paper comes from several directions, including low-dimensional topology, enumerative geometry, physics of BPS states, and three-dimensional cousins of elliptic genera with interesting modular properties.

In low-dimensional topology, we want to answer the~following question.
\begin{qn}
    Is there an~analogue of the~HOMFLY-PT polynomial for 3-manifolds?
\label{qn:3manHOMFLY}
\end{qn}
\noindent 
Among knot invariants, the~HOMFLY-PT polynomial plays a~special role as it unifies $A_{N-1}$~quantum group invariants of (super-)rank $N-1$ for all values of $N$. For example, the~Alexander polynomial, the~Jones polynomial, and its higher-rank cousins are all specialisations of the~HOMFLY-PT polynomial at $a=q^N$. Recent developments in 3d-3d correspondence suggest the~existence of $q$-series invariants of 3-manifolds that play a~role similar to that of the~Jones polynomial for knots. Then, a~natural question is whether these new invariants exhibit regularity and stabilisation with respect to rank.

To explain our motivation from the~perspective of enumerative invariants, it helps to recall different types of such invariants, summarised in Table \ref{tab:enumerative}.
The revolution in enumerative geometry starts with the~Gromov-Witten invariants that ``count'' stable maps $\phi: \Sigma_g \to X$ from a~Riemann surface of genus $g$ to the~target manifold $X$, which for our discussion we assume to be a~Calabi-Yau 3-fold. Topologically, such maps are classified by the~genus $g$ of $\Sigma_g$ and by the~homology class of its image, $\beta := \phi_* [\Sigma_g] \in H_2 (X,\mathbb{Z})$. Then, the~Gromov-Witten invariants of $X$ are defined in terms of the~intersection theory on the~moduli space of stable maps, $\mathcal{M}_g (X,\beta)$, with fixed $g$ and $\beta$. Although the~resulting numerical invariants $\text{GW}_g (X,\beta)$ have the~interpretation of ``counting'' stable maps, they are often rational rather than integer, because of denominators that account for automorphisms and resulting multi-valued perturbations.
\begin{table}[h]
	\begin{centering}
		\begin{tabular}{|c||c|c|c|c|}
			\hline
			~$\phantom{\int^{\int^\int}} ~ \phantom{\int_{\int}}$~ & ~Rational ($\hbar$)~ & ~Rational ($q$)~ &~Integer~ & ~Refined~ \tabularnewline
			\hline
			\hline
			$\phantom{\int} 
			\text{Closed} \phantom{\int_{\int}}$ &
			\begin{tabular}{c}
				GW \\
				(stable maps)
			\end{tabular} &
			\begin{tabular}{c}
				Bare curves \\
				(constants not\\ perturbed)
			\end{tabular} &
		\begin{tabular}{c}
			DT/GV \\
			(ideal sheaves)
		\end{tabular} & Equivariant
			\tabularnewline
			\hline
			$\phantom{\int} 
			\text{Open} \phantom{\int_{\int}}$ &
			\begin{tabular}{c}
				Open GW \\
				(relative stable\\ maps)
			\end{tabular} &
			\begin{tabular}{c}
				Bare curves, \\
				boundary in skein
			\end{tabular}
			& BPS invariants & \begin{tabular}{c}
				Knot \& 3-mfld \\
				homologies
			\end{tabular}
			\tabularnewline
			\hline
		\end{tabular}
		\par\end{centering}
		\medskip
	\caption{\label{tab:enumerative} Enumerative invariants.}
\end{table}

We emphasize the~non-integrality of Gromov-Witten invariants to help in comparing with many integer-valued enumerative invariants of $X$ that play an~important role in recent developments. The~prominent examples are the~Gopakumar-Vafa and Donaldson-Thomas invariants, which are close cousins 
%of each other 
and were independently discovered around the~same time~\cite{DT,GV}. Much like Donaldson invariants of 4-manifolds \cite{Don}, the~Donaldson-Thomas invariants of $X$ were originally formulated via analysis of six-dimensional gauge theory on $X$. In modern literature, one often uses an~equivalent formulation in terms of algebraic geometry of ideal sheaves $\mathcal{I}_{Z}$ of subschemes $Z \subset X$, such that $\chi (O_Z) = n$ and $[Z] = \beta \in H_2 (X, \mathbb{Z})$. In particular, Donaldson-Thomas invariants of $X$ are labelled basically by the~same data as Gromov-Witten invariants (except that the~genus $g$ is replaced by the~Euler characteristic~$n$) and defined similarly, via integration over the~virtual fundamental class of the~moduli space. The~resulting numerical invariants take integer values,  $\text{DT}_n (X,\beta) \in \mathbb{Z}$, and, physically, have an~interpretation as graded Euler characteristics of the~spaces of BPS states, $\mathcal{H}^{\text{BPS}}_{*,n,\beta} (X)$.
Conjecturally, these integer-valued invariants are connected with non-integer Gromov-Witten invariants via a~universal relation that takes the~same form for all~$X$. Schematically, it looks like \cite{MNOP,GV,Katz}
\begin{equation}
\xymatrixcolsep{9pc}\xymatrix{
\sum\limits_n \text{DT}_n \, q^n \quad \ar@/^/[r]^{\text{$q = e^{\hbar} \to 1$}} &
\quad \exp \, \left( \sum\limits_m \text{GW}_m \, \hbar^m \right) \ar@/^/[l]^{\text{Borel resum}}}
\label{DTviaGW}
\end{equation}
and involves two exponentials.
One exponential appears on the~right-hand side, replacing the~formal series $\sum_m \text{GW}_m \hbar^m$ -- which in general has zero radius of convergence and is often called  Gromov-Witten potential -- by its exponential. The~second relates the~variables on the~two sides, $q = e^{\hbar}$, and is one of the~keys to integrality. As indicated in the~third column of Table \ref{tab:enumerative}, there is a~recently discovered \cite{ES} middle ground: counts of so-called bare curves, i.e.~holomorphic curves with symplectic area zero components left unperturbed. Here the~count is naturally in terms of $q=e^\hbar$ rather than $\hbar$ itself, in a~sense corresponding to the~contributon to Gromov-Witten invariants of degree one in the~Goupakumar-Vafa formula. Bare curves are also key to the~understanding of open curve counts: boundaries of curves give line defects in Chern-Simons theory and open bare curves should be counted by the~values of their boundary in the~skein module of the~brane where they end.  

It is natural to ask whether the~space $\mathcal{H}^{\text{BPS}}_{*,n,\beta} (X)$ itself contains more information than its Euler characteristic that yields $\text{DT}_n (X,\beta)$. This leads to the~notion of {\it refined} BPS invariants, which in general can ``jump'' as one varies the~complex structure of $X$. However, when $X$ is rigid (e.g. toric), such refined invariants are well defined and indeed provide a~more detailed information than Donaldson-Thomas (DT) or Gopakumar-Vafa (GV) invariants~\cite{KS1}.

All enumerative invariants described so far can have an~{\it open} analogue, which involves the~data of the~Calabi-Yau 3-fold $X$ together with a~Lagrangian submanifold $L \subset X$. The~open Gromov-Witten invariants of $(X,L)$ are then defined as count of (generalised) stable maps from {\it bordered} surfaces $\Sigma$, such that the~boundary of $\Sigma$ lands on $L$ (for early work see e.g.~\cite{KL,GZ,LS}). Just like closed GW invariants, their open cousins are $\mathbb{Q}$-valued and, based on physics predictions, should satisfy \eqref{DTviaGW} with suitably defined integer {\it open}~DT~invariants. Unfortunately, the~theory of open DT invariants and their refinements is not developed at present, except in a~few special cases.

When $X$ is toric and $L \cong S^1 \times \mathbb{R}^2$ is compatible with the~torus action, one can compute (refined) open DT~invariants of $(X,L)$ via counting (skew) 3d partitions \cite{AKMV,ORV,IKV}. Another class of examples where an~easily computable expression for open DT invariants was recently proposed involves $(X,L)$ labelled by decorated graphs, the~so-called plumbing graphs~\cite{GPPV}. In yet another class of examples, related to knots, it was argued that refinement of open DT~invariants is equivalent to the~data of homological knot invariants~\cite{GSV}. So, in principle, if one knows the~latter, it can be used as a~definition of refined open DT invariants at least in these special cases, until a~better more universal definition is found.

Therefore, from the~viewpoint of enumerative geometry, a~challenge is to produce new families of examples where open DT invariants can be easily computed, either via combinatorial techniques, or via representation theory, that can hopefully shed light on the~general case.

\begin{qn}
Can we find new easy-to-work-with definitions of open DT invariants, at least for special classes of $(X,L)$?
\label{qn:DT}
\end{qn}
%\medskip
%\noindent
%{\bf Problem:} {\it Find new easy-to-work-with definitions of open DT invariants, at least for special classes of} $(X,L)$.
%\medskip

\noindent
Our present work can be viewed as a~step toward addressing this problem, for a~particular choice of $X$ and $L$, that also makes contact with recent developments of~\cite{DE,ES}. In particular, following \cite{OV,GSV}, we consider the~enumerative geometry of HOMFLY-PT knot invariants, along with their refined/categorified analogues. We relate it to new $q$-series invariants of 3-manifolds mentioned earlier, thus presenting some evidence that Question~\ref{qn:3manHOMFLY} may have an~affirmative answer.
As pointed out in \cite[sec.2.9]{GPPV}, a~similar relation between $\widehat{Z}$-invariants and enumerative geometry at finite $N$ helps to understand the~origin of Spin$^c$ structures, which (non-canonically) can be identified with $b \in H_1 (L,\mathbb{Z})$ and play a~role similar to $\beta \in H_2 (X,\mathbb{Z})$ in the~closed case.
Moreover, in the~large-$N$ limit we find a~few more surprises that we summarise shortly in the~form of new conjectures.

In finite rank $N$, the~$q$-series invariants $\widehat{Z}$ provide a~non-perturbative definition of Chern-Simons theory with complex gauge group. They can also be viewed as $\mathcal{U}_q (\mathfrak{sl}_N)$ quantum group invariants for generic values of the~parameter $|q|<1$. Surprisingly, as a~function of~$q$, $\widehat{Z} = \sum_n \text{DT}_n q^n$ turns out to be either a~character of a~logarithmic vertex algebra, or a~Ramanujan mock modular theta function, or a~$q$-series with integer coefficients and more exotic modular properties\footnote{A simple class of 3-manifolds for which the~modular properties of $\widehat{Z} (M_3)$ have not yet been explicitly identified consists of surgeries on the~figure-8 knot \cite{GM}.} that are expected to be a~variant of quantum modularity~\cite{BMM1,CCFGH,BMM2,CFS}.
In order to study the~large-$N$ behaviour of these invariants, one needs powerful tools to compute them for any rank $N>1$. Work in this direction was recently initiated in \cite{Chung,Park1}.

Studying the~large-$N$ behaviour of $\widehat{Z} (Y)$ for general 3-manifold $Y$ is a~very interesting but challenging problem. (See \cite[sec.7]{GPV} for a~brief survey of partial results in this direction and tools that could potentially be useful.) %to tackle this problem.) 
In this paper, we focus our attention on a~simpler version of this problem by taking $Y$  to be a~knot complement $S^3 \setminus K$.
Following \cite{GM,Park1}, we denote
\begin{equation}
F_K^{SU(N)} (x_1, \ldots, x_{N-1} ,q) \; := \; \widehat{Z} \left( S^3 \setminus K \right),
\label{FKZhat}
\end{equation}
which depends on extra variables $(x_1, \ldots, x_{N-1})$ that take values in the~maximal torus of the~complexified group $G_{\mathbb{C}}=SL(N,\mathbb{C})$. Even in this special case, the~study of the~large-$N$ behaviour is highly nontrivial, and we hope to report on it in future work. For the~purpose of the~present paper, we specialise even further to the~case where only $x_1 \equiv q x \in \mathbb{C}^*$ is nontrivial and the~rest are $x_i = q$ for $i \ne 1$. This corresponds to the~choice of a~one-dimensional subspace in the~weight lattice of $G=SU(N)$ associated with symmetric representations. With this specialisation, which we denote by $F_K^{SU(N),sym}(x,q)$, we will be able to understand the~large-$N$ behaviour of \eqref{FKZhat}. %for reasons that will become clear shortly.
From the~Chern-Simons theory perspective, the~variable $x$ is a~holonomy eigenvalue along the~meridian of the~knot $K$ and is one of the~variables in the~$A$-polynomial \cite{Guk}. It can also be understood as a~variable in the~Alexander polynomial $\Delta_K (x)$ \cite{GM}. Both of these polynomials will play an~important role in our story. From the~modularity viewpoint, $x$ plays the~role of a~Jacobi-type variable.

%As we hope to illustrate in this paper, each extra variable offers new directions and new opportunities. For example, t
The~interpretation of $\widehat{Z}$ as non-perturbative definition of Chern-Simons theory with complex gauge group (or 3d-3d correspondence) predicts that $F_K^{SU(N),sym}(x,q)$ should obey $q$-difference equations produced by the~quantisation of character varieties. One of the~main results in this paper, stated below in a~form of more precise conjectures, is that both $F_K^{SU(N),sym}(x,q)$ and the~$q$-difference equations themselves exhibit regularity with respect to~$N$. Schematically,
\begin{equation}
\hat{A}_K^{SU(N)} (\hat{x},\hat{y},q) F_K^{SU(N),sym}(x,q) = 0
\qquad \Leftrightarrow \qquad
\hat{A}_K (\hat{x},\hat{y},a,q)F_K(x,a,q) = 0,
\label{qdifffinvsinf}
\end{equation}
where $a=q^N$, and $\hat{A}_K (\hat{x},\hat{y},a,q)$ is the~annihilator of coloured HOMFLY-PT invariants of~$K$ (the quantum $a$-deformed $A$-polynomial) \cite{AV,FGS}. In more detail, basing on the~examples studied in this paper, we propose the~following:

\begin{conj}[$a$-deformed $F_K$]\label{$a$-deformed $F_K$}
    For every knot $K\subset S^3$, there exists a~three-variable function $F_K(x,a,q)$ interpolating all the~$F_K^{SU(N),sym}$ in the~following sense: 

    \begin{align}
        \label{eq:specializetoSUN}
       & F_K(x,q^N,q)  = F_K^{SU(N),sym}(x,q),\\
       & \hat{A}_K(\hat{x},\hat{y},a,q)F_K(x,a,q) = 0.
    \end{align}

    Moreover, it has the~following properties:$\,$\footnote{Here we are using the~\emph{reduced} normalisation. For the~\emph{unreduced} normalisation, we should have, for instance, \[\lim_{q\rightarrow 1}F_K(x,q^N,q) = \qty(\frac{x^{1/2}-x^{-1/2}}{\Delta_K(x)})^{N-1},\] and \[F_K(x^{-1},a,q) = (-1)^{N-1}F_K(a^{-1}x,a,q).\]} 
    \begin{align}
    \label{Weyl_symmetry}
    F_K(x^{-1},a,q) & = F_K(a^{-1} x,a,q),\\
    \label{eq:Alexanderlimit}
    F_K(x,1,q) & = \Delta_K(x), \\
     F_K(x,q,q) & = 1,\\
    \label{eq:mixedlimit}
    \lim_{q\rightarrow 1}F_K(x,q^N,q) & = \frac{1}{\Delta_K(x)^{N-1}}.
    \end{align}
    Its asymptotic expansion should agree with that of the~coloured HOMFLY-PT polynomials. That is, 
    \begin{equation}\label{eq:asymptotics}
    \log F_K(e^{r\hbar},a,e^\hbar) = \log P_r(K;a,e^{\hbar})
    \end{equation}
    as $\hbar$-series. 
\end{conj}

\begin{conj}[$(a,t)$-deformed $F_K$]\label{$a,t$-deformed $F_K$}
    For every knot $K\subset S^3$, there exists a~four-variable function $F_K(x,a,q,t)$ such that
    \begin{align}
    & F_K(x,a,q,-1)  = F_K(x,a,q), \\
    & \hat{A}_K(\hat{x},\hat{y},a,q,t)F_K(x,a,q,t) = 0.
    \end{align}
    Its asymptotic expansion should agree with that of the~superpolynomials. That is, 
    \begin{equation}
    \log F_K(e^{r\hbar},a,e^\hbar,t) = \log \mathcal{P}_r(K;a,e^{\hbar},t)
    \end{equation}
    as $\hbar$-series. 
\end{conj}

In line with enumerative interpretations of the~HOMFLY-PT polynomial in terms of counts of open holomorphic curves with boundary on the~knot conormal in the~resolved conifold, we give a~similar enumerative interpretation of $F_{K}(a,q)$ in terms of counts of holomorphic curves with boundary on the~knot complement. In the~case of fibered knots, the~knot complement Lagrangian in $T^{\ast}S^{3}$ can be shifted off of the~zero section $S^{3}$ and then considered as a~Lagrangian in the~resolved conifold. Here the~interpretation of $F_{K}(y,a,q)$ as a~count of curves with boundary in homology class $\log y$ is directly analogous to the~knot conormal case (where the~homology variable is $\log x$), and the~classical limit ($q\to 1$ and $a\to 1$) was studied in \cite{DE}. For non-fibered knots the~situation is further complicated by the~appearance of intersection points between the~knot complement Lagrangian and the~zero section. Here, as in \cite{ES}, we apply Symplectic Field Theory (SFT) stretching which leaves cotangent fibers in $T^{\ast}S^{3}$. As we will discuss, these fibers are connected by Reeb chords at infinity that appear as negative ends in extra curves to be counted. For reasons of invariance of such counts, the~values assigned to the~Reeb chords are not arbitrary. They are functions of $(x,a,q)$ determined in the~semiclassical case by augmentations of a~differential graded algebra and in the~full quantum case by an~analogous Legendrian SFT equation, in analogy with \cite[Section~3]{EN}. Also, similar to \cite[Section 6]{EN}, the~operator equation $\hat{A}_K=0$ has an~interpretation in terms of curve counts at the~ideal boundary at infinity of the~knot complement Lagrangian in $T^\ast S^3$.        

Below we provide evidence for Conjectures \ref{$a$-deformed $F_K$} and \ref{$a,t$-deformed $F_K$} by showing that they hold true for $(2,2p+1)$ torus knots and for the~figure-eight knot. For example, for the~trefoil knot, $F_K$~%the~two-variable series \eqref{FKZhat} 
itself is a~``deformation'' of the~Dedekind eta-function\footnote{In this ``deformation'' the~$m$-th term is multiplied by $x^m-x^{-m}$. Another version, that appears e.g.\ in \cite{Park2}, is when the~$m$-th term is multiplied just by $x^m$; it gives a~genuine deformation of the~$\eta$-function, in which the~latter is recovered by taking $x \to 1$ and is related to the~invariant $F_K (x,q)$ by further anti-symmetrisation with respect to $x \to x^{-1}$.}
\begin{equation}
\eta (q)
= q^{\frac{1}{24}} \prod_{n=1}^{\infty} (1-q^n)
= \sum_{m=1}^{\infty} \epsilon_m q^{\frac{m^2}{24}}
\qquad \leadsto \qquad
F_{ 3_1} = \sum_{m=1}^{\infty} \epsilon_m q^{\frac{m^2}{24}} (x^m-x^{-m})
\end{equation}
and in this paper we propose a~further two-variable deformation/refinement \eqref{eq:t-deformed F_K for torus knots}, with two extra variables $a$ and $t$.
Assuming that functions $F_K (x,a,q)$ exist for every knot $K$ and inspired by \cite{Zag}, we may ask:
\begin{qn}
What are the~modular properties of $F_K (x,a,q)$? Is there a~number theory (or vertex algebra) interpretation of the~$q$-difference equations \eqref{qdifffinvsinf}?
\label{qn:FKmodular}
\end{qn}
\noindent
Hoping that future work will shed light on the~first part of the~question, in this paper we will only offer some clues regarding the~second part.
For example, one important lesson that follows from the~general framework reviewed in the~introduction and used in the~rest of this paper is that the~$q$-difference equation for the~mirror knot $m(K)$ is related to that of $K$ simply be replacing $q \mapsto q^{-1}$, $a \mapsto a^{-1}$ (and $t \mapsto t^{-1}$ in the~refined case).
Since $\hat{A}_K(\hat{x},\hat{y},a,q)$ is a~rational function of these variables, it transforms in a~simple way under $K \mapsto m(K)$.
On the~other hand, the~solutions to the~corresponding $q$-difference equation, $F_K(x,a,q)$ and $F_{m(K)} (x,a,q)$, are related in a~highly nontrivial way, generalising variants of quantum modularity found in \cite{BMM1,CCFGH,BMM2,CFS}. This is interesting also for the~interpretation of $F_K(x,a,q)$ and $F_{m(K)} (x,a,q)$ as characters of ``dual'' logarithmic vertex algebras, where it gives a~nice structural property shared by completely different vertex operator algebras.

The~rest of the~paper is organised as follows. In Section \ref{sec:Z/F} we introduce $F_K$ invariants ($\widehat{Z}$~invariants for knot complements). Their relations with low-dimensional topology, physics, and enumerative geometry are presented in Sections \ref{sec:HOMFLY and A} and \ref{sec:Enumerative geometry}. Section \ref{sec:a-deformed F_K} contains explicit results on $a$-deformed $F_K$ invariants for $(2,2p+1)$-torus knots and figure-eight knot, whereas section \ref{sec:t-deformation} proposes a~$t$-deformation. In Section \ref{sec:Quantum modularity} we show how the~analysis of the~behaviour of $F_K$ invariants under taking the~mirror of the~knot $K\to m(K)$ provides a~new area for studies of the~quantum modularity. Finally, in Section \ref{sec:Future directions} we discuss interesting problems for future research.

\section{\texorpdfstring{$\widehat{Z}$ and $F_K$}{Z/F} invariants}\label{sec:Z/F}
%\section{\texorpdfstring{$F_K$}{FK} invariants for SU(2) and SU(\texorpdfstring{$N$}{N})}

%A few years ago, 
In their study of $3$d $\mathcal{N}=2$ theory $T[Y]$ for 3-manifolds $Y$, Gukov-Putrov-Vafa \cite{GPV} and Gukov-Pei-Putrov-Vafa \cite{GPPV} conjectured the~existence of the~$3$-manifold invariants $\widehat{Z}(Y)$ (also known as ``homological blocks'' or ``GPPV invariants'') valued in $q$-series with integer coefficients.
These $q$-series invariants exhibit peculiar modular properties, the~exploration of which was initiated in~\cite{BMM1,CCFGH,BMM2,CFS}.

More recently, Gukov-Manolescu \cite{GM}
introduced a~version of $\widehat{Z}$ for knot complements, which they called $F_K$: if $K\subset S^3$ is a~knot, then $F_K=\widehat{Z}(S^3\setminus K)$.
The~motivation was to study $\widehat{Z}$ more systematically using Dehn surgery. 
%Before stating their conjecture, let us 
Recall that the~Melvin-Morton-Rozansky expansion \cite{MM,BNG,Roz1,Roz2} (also known as ``loop expansion'' or ``large colour expansion'') of the~coloured Jones polynomials is the~asymptotic expansion near $\hbar \rightarrow 0$ while keeping $x=q^r=e^{r\hbar}$ fixed:  
\begin{equation}
    J_r(K;q=e^\hbar) = \sum_{j\geq 0}\frac{p_j(x)}{\Delta_K(x)^{2j+1}}\frac{\hbar^j}{j!},\quad\quad p_j(x)\in \mathbb{Z}[x,x^{-1}],\quad p_0 = 1.
\end{equation}
Here $\Delta_K(x)$ is the~Alexander polynomial of $K$. 
%Now we can state 
The main conjecture of \cite{GM} was then the~following. 
\begin{conj}\label{conjGukovManolescu}
    For every knot $K\subset S^3$, there exists a~two-variable series 
    \begin{equation}\label{eq:GMconj}
    F_K(x,q) = \frac{1}{2}\sum_{\substack{m\geq 1\\m\text{ odd}}} f_m(q) (x^{m/2}-x^{-m/2}),\quad\quad f_m(q)\in \mathbb{Z}[q^{-1},q]]
    \end{equation}
    such that its asymptotic expansion agrees with the~Melvin-Morton-Rozansky expansion of the~coloured Jones polynomials\footnote{\cite{GM}  uses the~unreduced normalisation. In the~reduced normalisation, used in the~major part of this paper, \eqref{eq:GM-Melvin-Morton} reads $F_K(x,q=e^{\hbar}) = \sum_{j\geq 0}\frac{p_j(x)}{\Delta_K(x)^{2j+1}}\frac{\hbar^j}{j!}.$}: 
    \begin{equation}\label{eq:GM-Melvin-Morton}
        F_K(x,q=e^{\hbar}) = (x^{1/2}-x^{-1/2})\sum_{j\geq 0}\frac{p_j(x)}{\Delta_K(x)^{2j+1}}\frac{\hbar^j}{j!}.
    \end{equation}
    Moreover, this series is annihilated by the~quantum $A$-polynomial: 
    \begin{equation}
        \hat{A}_K(\hat{x},\hat{y},q)F_K(x,q) = 0.
    \end{equation}
\end{conj}
Conjecture \ref{conjGukovManolescu} concerns $G=SU(2)$. An~extension 
%of this conjecture 
to arbitrary $G$ was studied in \cite{Park1}. In particular, the~existence of an~$SU(N)$ generalisation of $F_K$, denoted~$F_K^{SU(N)}$, was conjectured and it was observed that its specialisation to symmetric representations, $F_K^{SU(N),sym}$, is annihilated by the~corresponding quantum $A$-polynomial: 
\begin{equation}
    \hat{A}_K(\hat{x},\hat{y},a=q^N,q)F_K^{SU(N),sym}(x,q) = 0.
\end{equation}
%A natural question arising from this observation is: 
%\begin{qn}
%    Is there an~$a$-deformed (i.e. HOMFLY-PT analogue of) $F_K$?
%\end{qn}
From this perspective it is natural to ask about the~large-$N$ behaviour and existence of an~$a$-deformed (i.e. HOMFLY-PT analogue of) the~$F_K$ invariant. This was the~starting point of this paper and, as we will see, HOMFLY-PT analogues of $F_K$
%that there are 
exist indeed.
%for various examples of knot complements. 

\section{Relations with low-dimensional topology and physics}\label{sec:HOMFLY and A}
%\section{Knot invariants%: HOMFLY-PT and $A$-polynomials
%}

In this section we discuss the~connection between %the invariant
$F_K$ and other knot invariants, such as HOMFLY-PT and $A$-polynomials, and 3d $\mathcal{N}=2$ effective theories engineered on the~worldsheets of M5-branes.
%corresponding to knot complements.

\subsection{HOMFLY-PT polynomials}\label{sec:HOMFLY}
If $K\subset S^{3}$ is a~knot, then its HOMFLY-PT polynomial $P(K;a,q)$  is a~topological invariant \cite{HOMFLY,PT} which can be calculated via the~skein relation:
\begin{equation}
    a^{1/2}P(\overcross)-a^{-1/2}P(\undercross)=(q^{1/2}-q^{-1/2})P(\nocross)
\end{equation}
with a~normalisation condition $P(0_{1};a,q)=1$. It is called {\it reduced} normalisation and corresponds to dividing by the~full natural HOMFLY-PT polynomial for the~unknot (here denoted by bar):
\begin{equation}\label{eq:reduced vs unreduced}
\begin{split}
    P(K;a,q)&=\frac{\bar{P}(K;a,q)}{\bar{P}(0_{1};a,q)},\\
    \bar{P}(0_{1};a,q)&=\frac{a^{1/2}-a^{-1/2}}{q^{1/2}-q^{-1/2}}.
\end{split}
\end{equation}
We use the~reduced normalisation in the~major part of the~paper, except %the~curve counts in Sections \ref{sec:Enumerative geometry} and \ref{sec:Quantum modularity}, as well as the~discussion of the~unknot in Sections \ref{sec:a-deformed F_K} and \ref{sec:t-deformation}. There we use the~unreduced normalisation (denoted by bar) which is related to the~reduced one by
the~geometric considerations in Sections \ref{sec:Enumerative geometry} and \ref{sec:Quantum modularity}, where we analyse curve counts leading to fully unreduced normalisation -- corresponding to $\bar{P}(K;a,q)$ -- and explain how to obtain the~reduced one.

More generally, the~coloured HOMFLY-PT polynomials $P_{R}(K;a,q)$ are similar polynomial knot invariants depending also on a~representation $R$ of $SU(N)$. In this setting, the~original HOMFLY-PT corresponds to the~defining representation. We will be interested mainly in HOMFLY-PT polynomials coloured by the~totally symmetric representations $S^{r}$ with $r$~boxes in one row of the~Young diagram. In order to simplify the~notation, we will denote them by $P_{r}(K;a,q)$ and call %them simply 
HOMFLY-PT polynomials.

From our perspective, the~most important property of HOMFLY-PT polynomials is the~fact that after the~substitution $a=q^N$, they reduce to the~$SU(N)$ coloured Jones polynomials: 
\begin{equation}
    P_r (K;q^N,q)=J^{SU(N)}_r (K;q),
\end{equation}
whose asymptotic expansion (as a~series in $\hbar$) agrees with symmetric $SU(N)$ $F_K$ invariants \cite{Park1,GM}. This connection can be considered as the~base of the~relation between HOMFLY-PT polynomials and $a$-deformed $F_K$ invariants.

In \cite{DGR,GS1}, a~$t$-deformation of HOMFLY-PT polynomial was proposed. The~superpolynomial $\mathcal{P}_r(K,a,q,t)$ was defined as a~Poincar\'e polynomial of the~triply-graded homology that categorifies the~HOMFLY-PT polynomial:
\begin{equation}
\begin{split}
    P_{r}(K;a,q)&=\sum_{i,j,k}(-1)^{k}a^{i}q^{j}\dim \mathcal{H}^{S^{r}}_{i,j,k}(K),\\   
   \mathcal{P}_{r}(K;a,q,t)&=\sum_{i,j,k}a^{i}q^{j}t^{k}\dim \mathcal{H}^{S^{r}}_{i,j,k}(K).
\end{split}   \label{Pr}
\end{equation}
We will use this categorification to propose an~$(a,t)$-deformed $F_K$ invariant in Section \ref{sec:t-deformation}.

\subsection{\texorpdfstring{$A$}{A}-polynomials}

The $A$-polynomial is a~knot invariant associated to a~character variety of a~complement of a~given knot $K$ in $S^3$ \cite{CCGLS}. It takes the~form of an~algebraic curve $A_K(x,y)=0$, for $x,y\in\mathbb{C}^*$. According to the~volume conjecture, it also captures the~asymptotics of coloured Jones polynomials $J_r(K;q)$ for large colours $r$. The~quantisation of the~$A$-polynomial encodes information about all colours, not only large. Namely, it gives the~recursion relations satisfied by $J_r(K;q)$, which can be written in the~form
 \begin{equation}
     \hat{A}_K(\hat x,\hat y) J_{*}(K;q) = 0,
 \end{equation}
where $\hat x, \hat y$ act by
\begin{equation}
    \hat{x} J_r = q^r J_r, \quad \quad \hat{y}J_r = J_{r+1},
\end{equation}
and satisfy the~relation $\hat{y}\hat{x}=q\hat{x}\hat{y}$. The~above conjecture was proposed independently in the~context of quantisation of Chern-Simons theory \cite{Guk} and in parallel mathematics developments~\cite{Gar}. The~operator $\hat{A}_K(\hat x,\hat y)$ is referred to as the~quantum $A$-polynomial; in the~limit $q=1$ it reduces to the~polynomial defining the~$A$-polynomial algebraic curve~$A_K(x,y)$.

The above conjectures were generalised to coloured HOMFLY-PT polynomials \cite{AV} and coloured superpolynomials \cite{FGSA,FGS}, which we introduced in (\ref{Pr}). In these cases the~objects mentioned in the~previous paragraph become $a$- and $t$-dependent. In particular, the~asymptotics of coloured superpolynomials $\mathcal{P}_{r}(K;a,q,t)$ for large $r$ is captured by an~algebraic curve called super-$A$-polynomial, defined by an~equation $A_K(x,y,a,t)=0$. For $t=-1$ it reduces to $a$-deformed $A$-polynomial, and upon setting in addition $a=1$ we obtain the~original $A$-polynomial as a~factor. For brevity, all these objects are often referred to as~$A$-polynomials. A~quantisation of the~super-$A$-polynomial gives rise to quantum super-$A$-polynomial $\hat{A}_K(\hat x,\hat y,a,q,t)$, which is an~operator that imposes recursion relations for coloured superpolynomials:
 \begin{equation}
     \hat{A}_K(\hat x,\hat y,a,q,t) \mathcal{P}_{*}(K;a,q,t) = 0.
 \end{equation}
 A~universal framework that enables to determine a~quantum $A$-polynomial from an~underlying classical curve $A(x,y)=0$ was proposed in \cite{GS2} (irrespective of extra parameters these curves depend on, and also beyond examples related to knots).
 
The quantum $A$-polynomial $\hat{A}_K(\hat x,\hat y)$ can be regarded as a~polynomial in $\hat{y}$, whose coefficients depend on $q$ and $x=q^r$. It was conjectured in \cite{GM} that at the~same time, the~quantum $A$-polynomial is an~operator that annihilates $F_K$, once $\hat{x}$ is interpreted as a~multiplication by $x$ and $\hat{y}$ acts by $\hat{y} F_K(x, q) = F_K(qx, q)$. While the~same form of quantum $A$-polynomial $\hat{A}_K(\hat x,\hat y)$ arises in the~analysis of coloured Jones polynomial and $F_K$ invariants, there is a~subtle but important difference between these two situations, which has to do with the~initial conditions that need to be imposed.

One of the~main results of this paper is the~statement that ($a,t$)-deformed $F_K$ invariants are annihilated by quantum super-$A$-polynomial $\hat{A}_K(\hat x,\hat y,a,q,t)$ that we presented above (or its $t=-1$ limit in case of $a$-deformed $F_K$ invariants). We verify this statement for the~family of $(2,2p+1)$ torus knots and figure-eight knot. Apart from the~conceptual importance, this statement implies that we can simply take advantage of expressions for quantum super-$A$-polynomials derived before, e.g. in \cite{FGS,FGSS}. Nonetheless, to determine $F_K$ using these quantum $A$-polynomials -- or to check that they are indeed annihilated by~$\hat{A}_K(\hat x,\hat y,a,q,t)$ -- we need to use proper initial conditions. Furthermore, our conventions in this paper are such that, in comparison with \cite{FGS}, we need to rescale $\hat{x}$ by $q$ and $\hat{y}$ by $a/q$. We discuss all these issues in detail in the~following sections.

\subsection{3d-3d correspondence}\label{sec:Physical}
%\subsection{Physical interpretations}

From the~physical point of view, the~$\widehat{Z}$-invariants of a~3-manifold $Y$ encode the~BPS spectrum of $N$ fivebranes supported on $\mathbb{R}^2 \times S^1 \times Y$, where $Y$ is embedded (as a~zero-section) inside the~Calabi-Yau space $T^* Y$ and $\mathbb{R}^2 \times S^1 \subset \mathbb{R}^4 \times S^1$:
\begin{equation}
\begin{split}\text{space-time}:\quad & \mathbb{R}^{4}\times S^{1}\times T^* Y\\
 & \cup\phantom{\ \times S^{1}\times\ \ }\cup\\
N~\text{M5-branes}:\quad & \mathbb{R}^{2}\times S^{1}\times Y.
\end{split}
\label{MgeneralY}
\end{equation}
Taking the~large-$N$ limit of this system for general $Y$ is highly nontrivial (see \cite[sec.7]{GPV} and \cite[Remark 2.4]{ES}). However, when $Y$ is a~knot complement $M_{K}:=S^{3}\backslash K$
then there is an~equivalent description of the~physical system \eqref{MgeneralY} for which the~study of large-$N$ behaviour can be reduced to the~celebrated ``large-$N$ transition'' \cite{GV,OV}.

Note that from the~viewpoint of 3d-3d correspondence, $N$ fivebranes on $Y = M_{K}$ produce a~4d $\mathcal{N}=4$ theory -- which is a~close cousin of 4d $\mathcal{N}=4$ SYM but is {\it not} 4d $\mathcal{N}=4$ SYM -- on a~half-space $\mathbb{R}^3 \times \mathbb{R}_+$ coupled to 3d $\mathcal{N}=2$ theory $T[M_K]$ on the~boundary.
Indeed, near the~boundary $T^2 =\Lambda_K= \partial M_K$, the~compactification of $N$ fivebranes produces a~4d $\mathcal{N}=4$ theory which has moduli space of vacua $\text{Sym}^N (\mathbb{C}^2 \times \mathbb{C}^*)$ \cite{CGPS}.
(Recall that the~moduli space of vacua in 4d $\mathcal{N}=4$ super-Yang-Mills is $\text{Sym}^N (\mathbb{C}^3)$.)
The $SU(N)$ gauge symmetry of this theory appears as a~global symmetry of the~3d boundary theory $T[M_K]$. In particular, the~variables $x_i \in \mathbb{C}^*$ in \eqref{FKZhat} are complexified fugacities for this global (``flavour'') symmetry.
For $G=SU(2)$, the~moduli space of vacua of the~knot complement theory $T[M_K]$ gives precisely the~$A$-polynomial of $K$. And, similarly, $G_{\mathbb{C}}$ character varieties of $M_{K}$ are realised as spaces of vacua in $T[M_{K},SU(N)]$ with $G=SU(N)$ \cite{FGS,FGSS}.

Now, as promised, let us give another, equivalent description of the~physical system \eqref{MgeneralY} with $Y = M_{K}$, where the~large-$N$ behaviour is easier to analyse:
\begin{equation}
\begin{split}\text{space-time}:\quad & \mathbb{R}^{4}\times S^{1}\times T^* S^3\\
 & \cup\phantom{\ \times S^{1}\times\ \ }\cup\\
N~\text{M5-branes}:\quad & \mathbb{R}^{2}\times S^{1}\times S^3 \\
\rho~\text{M5-branes}:\quad & \mathbb{R}^{2}\times S^{1}\times L_K.
\end{split}
\label{Mdeformed}
\end{equation}
This brane configuration is basically a~variant of \eqref{MgeneralY} with $Y=S^3$ and $\rho$ extra M5-branes supported on $\mathbb{R}^2 \times S^1 \times L_K$, where $L_K \subset T^* S^3$ is the~conormal bundle of the~knot $K \subset S^3$. There is, however, a~crucial difference between fivebranes on $S^3$ and $L_K$. Since the~latter are non-compact in two directions orthogonal to $K$, they carry no dynamical degrees of freedom away from $K$. One can path integrate those degrees of freedom along $K$; this effectively removes $K$ from $S^3$ and puts the~corresponding boundary conditions on the~boundary $T^2 = \partial M_K$. The~resulting system is precisely \eqref{MgeneralY} with $Y = M_{K}$. Equivalently, one can use the~topological invariance along $S^3$ to move the~tubular neighbourhood of $K \subset S^3$ to ``infinity.'' This creates a~long neck $\cong \mathbb{R} \times T^2$ as in the~above discussion. Either way, we end up with a~system of $N$ fivebranes on the~knot complement and no extra branes on $L_K$, so that the~choice of $GL(\rho,\mathbb{C})$ flat connection on $L_K$ is now encoded in the~boundary condition for $SL(N,\mathbb{C})$ connection\footnote{To be more precise, it is a~$GL(N,\mathbb{C})$ connection, but the~dynamics of the~$GL(1,\mathbb{C})$ sector is different from that of the~$SL(N,\mathbb{C})$ sector and can be decoupled.} on $T^2 = \partial M_K$. In particular, the~latter has at most $\rho$ nontrivial parameters $x_i \in \mathbb{C}^*$, $i=1,\ldots, \rho$.

Although the~relation between $N$ fivebranes on a~knot complement and \eqref{Mdeformed} holds for any value of $\rho$ (with a~suitable identification of boundary conditions, of course), the~extreme values are somewhat special. The~maximal value $\rho=N$ is what one needs to study the~full-fledged $\widehat{Z}$-invariants,  cf. \eqref{FKZhat}. However, this, or any other choice of $r \sim O (N)$ make the~study of $N \to \infty$ rather challenging since both sets of fivebranes in \eqref{Mdeformed} need to be replaced by geometry and such generalised ``geometric transitions'' are not known. The~other extreme is when $\rho \sim O(1)$ as $N \to \infty$; in particular, in this paper we consider the~simplest such option $\rho=1$. In that case we can use the~geometric transition of Gopakumar and Vafa \cite{GV}, upon which there is one brane on $L_K$ and $N$ fivebranes on the~zero-section of $T^* S^3$ disappear. Then the~Calabi-Yau space $T^* S^3$ undergoes a~topology changing transition to $X$, the~total space of $\mathcal{O} (-1) \oplus \mathcal{O} (-1) \to \mathbb{C}{\bf P}^1$, the~so-called ``resolved conifold.''
Only the~Ooguri-Vafa fivebranes supported on the~conormal bundle $L_K$ remain:
\begin{equation}
\begin{split}\text{space-time}:\quad & \mathbb{R}^{4}\times S^{1}\times X\\
 & \cup\phantom{\ \times S^{1}\times\ \ }\cup\\
\rho~\text{M5-branes}:\quad & \mathbb{R}^{2}\times S^{1}\times L_{K}.
\end{split}
\label{Mresolved}
\end{equation}
Note that on the~resolved conifold side, i.e. after the~geometric transition, $\log a~= \text{Vol} (\mathbb{C}{\bf P}^1) + i \int B = N\hbar$ is the~complexified K\"ahler parameter which, as usual, enters the~generating function of enumerative invariants presented in Table~\ref{tab:enumerative}.

To summarise, a~system of $N$ fivebranes on a~knot complement \eqref{MgeneralY} is equivalent to a~brane system \eqref{Mresolved}, with a~suitable map that relates the~boundary conditions in the~two cases. There is another system, closely related to \eqref{Mresolved}, that one can obtained from \eqref{Mdeformed} by first reconnecting $\rho$ branes on $L_K$ with $\rho$ branes on $S^3$. This give $\rho$ branes on $M_K$ (that go off to infinity just like $L_K$ does) plus $N-\rho$ branes on $S^3$. Assuming that $\rho \sim O(1)$ as $N \to \infty$ (e.g. $\rho=1$ in the~context of this paper), after the~geometric transition we end up with a~system like \eqref{Mresolved}, except $L_K$ is replaced by $M_K$ and $\text{Vol} (\mathbb{C}{\bf P}^1) + i \int B = (N-\rho) \hbar$.
Both of these systems on the~resolved side compute the~HOMFLY-PT polynomials of $K$ coloured by Young diagrams with at most $\rho$ rows.

The leading, genus-0 contribution to the~generating function of enumerative invariants is the~twisted superpotential. It can be computed either on the~resolved side of the~transition, where $a$ is a~K\"ahler parameter, or on the~original (``deformed'') side, for a~family of theories labelled by $N$. Either way, one finds that the~twisted superpotential is given by the~double-scaling limit that combines large-colour and semiclassical limits of the~HOMFLY-PT polynomials \cite{FGS,FGSS}:
\begin{equation}
P_{r}(K;a,q)\stackrel[\hbar\rightarrow0]{r\rightarrow\infty}{\longrightarrow}\int\prod_{i}\frac{dz_{i}}{z_{i}}\exp\left[\frac{1}{\hbar}\widetilde{\mathcal{W}}(z_{i},x,a)+O(\hbar^{0})\right],
\end{equation}
with $x=q^{r}$ kept fixed. We can read off the~structure of $T[M_{K}]$
from the~terms in $\widetilde{\mathcal{W}}(z_{i},x,a)$:
\begin{equation}
\begin{split}\textrm{Li}_{2}\ensuremath{\left(a^{n_{Q}}x^{n_{M}}z_{i}^{n_{i}}\right)}\qquad & \longleftrightarrow\qquad\text{(chiral field)}\,,\\
\frac{\kappa_{ij}}{2}\log\zeta_{i}\cdot\log\zeta_{j}\qquad & \longleftrightarrow\qquad\text{(Chern-Simons coupling)}\,.
\end{split}
\label{eq:Li2 and logs dictionary}
\end{equation}
Each dilogarithm is interpreted as the~one-loop contribution of a~chiral superfield with charges $(n_{Q},n_{M},n_{i})$ under the~global symmetries $U(1)_{Q}$ (arising from the~internal 2-cycle in $X$) and~$U(1)_{M}$ (corresponding to the~non-dynamical gauge field on the~M5-brane), and the~gauge group $U(1)\times\ldots\times U(1)$. Quadratic-logarithmic terms are identified with Chern-Simons couplings among the~various $U(1)$ symmetries, with $\zeta_{i}$ denoting the~respective fugacities.

We can integrate out the~dynamical fields (whose VEVs are given by $\log z_{i}$) using the~saddle point approximation to obtain the~effective twisted superpotential:
\begin{equation}
\widetilde{\mathcal{W}}_{\textrm{eff}}(x,a)=\frac{\partial\widetilde{\mathcal{W}}(z_{i},x,a)}{\partial\log z_{i}}.
\end{equation}
Then, after introducing the~dual variable $y$ (the effective Fayet-Iliopoulos parameter), we arrive at the~$A$-polynomial:
%I think that we can write that we drop the~a-deformed.
\begin{equation}
\log y=\frac{\partial\widetilde{\mathcal{W}}_{\textrm{eff}}(x,a)}{\partial\log x}\qquad\Leftrightarrow\qquad A_K(x,y,a)=0.
\end{equation}

\section{Relations with enumerative geometry}
\label{sec:Enumerative geometry}
%\section{Curve counts on the~knot complement}
In this section we look at $F_K$ invariants and Alexander polynomials 
%appearing in the~Melvin-Morton-Rozansky expansion 
from the~point of view of the~enumerative geometry in the~spirit of large $N$ duality. For comparison with invariants already discussed, we point out that 
%although most of the~statemets of this section are independent of the~normalisation, let us stress that 
counts of holomorphic curves naturally give invariants in the~{\it fully unreduced} normalisation corresponding to $\bar{P}(K;a,q)$, see Sections \ref{sec:HOMFLY} and \ref{sec:a-deformed unknot results}. To get results in the~reduced normalisation, one divides by the~curve count (or equivalently unreduced invariant) of the~unknot. 

\subsection{Curve counts}
We start from the~deformed conifold $T^{\ast}S^{3}$ and two Lagrangians: the~knot conormal $L_K$ and the~knot complement $M_K$, which both have the~Legendrian conormal $\Lambda_{K}\subset ST^{\ast} S^{3}$ as ideal boundary. 
%Let $K\subset S^{3}$ be a~knot. Write $L_{K}\subset T^{\ast}S^{3}$ for its conormal and $M_{K}\subset T^{\ast} S^{3}$ for the~exact Lagrangian corresponding to the~knot complement. Then both have the~Legendrian conormal $\Lambda_{K}\subset ST^{\ast} S^{3}$ as ideal boundary. 
We shift $L_{K}$ off of the~zero-section $S^3$ along the~closed non-vanishing form $d\theta$ that generates $H^{1}(L_{K})= H^{1}(S^{1}\times\R^{2})$. We shift $M_{K}$ similarly along a~closed form $\beta$ that generates $H^{1}(M_{K})=\R$. We take this form to agree with the~form $d\mu$ that is dual to the~meridian circle on the~boundary of a~tubular neighbourhood of the~knot. If $K$ is fibered, then we can find a~non-zero such form $\beta$, otherwise not. 

We want to count (generalised) holomorphic curves with boundary on $L_{K}$ or $M_{K}$. There are two ways to do this for $L_{K}$, either we can consider $L_{K}$ as a~Lagrangian submanifold in the~resolved conifold $X$ or we can use a~sufficiently SFT-stretched almost complex structure on $T^{\ast}S^{3}$ for which all curves leave a~neighbourhood of the~zero section, see \cite[Section 2.5]{ES}. The~resulting counts (and in fact the~curves) are the~same. For $M_{K}$ the~second approach still works: after stretching $M_{K}$ intersects a~neighbourhood of $S^{3}$ in $T^{\ast} S^{3}$ in a~finite collection of cotangent fibers. Then, possible curves in the~inside region (near $S^3$) have boundaries on these fibers and positive punctures at Reeb chords corresponding to geodesics connecting them. The~dimension of such a~curve is 
\[ 
\dim = \sum_{j}(\ind(\gamma_{j})+1)\ge 2, 
\]  
where the~sum runs over positive punctures of the~curve, $\gamma_{j}$ is the~Reeb chord at the~puncture, and $\ind(\gamma_{j})$ the~Morse index of the~corresponding geodesic.
It follows that no such curve can appear after stretching, since the~outside part would then have negative index. This means that there is a~curve count also for $M_{K}$. As we will discuss below, although this curve count is well defined and invariant, when intersections between $S^{3}$ and $M_{K}$ cannot be removed, it is only one point in a~space of curve counts that also takes into account certain punctured curves. The~present discussion applies to the~more involved curve counts of punctured curves as well.   

Logarithms of the~variables $x$ and $y$ correspond to cycles in $\Lambda_{K}$: $\log x$ corresponds to the~meridian and $\log y$ to the~longitude.
Then $\log x$ is homologous to zero in~$L_{K}$ and generates~$H_{1}(M_{K})$. On the~other hand, $\log y$ is homologous to zero in~$M_{K}$ and generates~$H_{1}(L_{K})$. We write  
\begin{equation}
   \Psi_{K}(y,a,g_{s})=\sum_{r\ge 0} \bar{P}_{r}(K;a,q=e^{g_{s}}) y^{-r}=\exp(p_{K}(y,a,g_{s}))
\end{equation}
for the~generating function counting disconnected generalised holomorphic curves on $L_{K}$. Here a~curve that goes $r$ times around $\log y$ contributes $w a^{\deg}g_{s}^{-\chi}$ to the~coefficient of $y^{-r}$, where $g_s$ is the~string coupling constant, $\chi$ is the~Euler characteristic of the~curve, $w$ its rational weight, and $\deg$ its degree in~$H_{2}(T^{\ast}S^{3}\setminus S^{3})$ 
%Here there was $H_{2}(T^{\ast}DS^{3}\setminus S^{3})$ and I interpreted D as a~typo.
after capping off. The~logarithm $p_K(y,a,g_{s})$ is the~corresponding count of connected curves and we have
\begin{equation}
\label{eq:Psi_connected_expansion}
p_{K}(y,a,g_{s}) = g_{s}^{-1}W_{K}(y,a) + W_{K}^{0}(y,a)+g_{s} W_{K}^{1}(y,a)+\dots,
\end{equation}
where $W_{K}$ is the~disk potential and $W_{K}^{k}$ counts curves of Euler characteristic $\chi = -k$.

For $M_{K}$ we have analogously the~generating function of disconnected curves:
\begin{equation} \label{eq:Phi_full_expansion}
\Phi_{K}(x,a,g_{s})=\sum_{r\ge 0} V_{r}(a,e^{g_{s}})x^{r} = \exp(v_K(x,a,g_{s})),
\end{equation} 
where
\begin{equation} \label{eq:Phi_connected_expansion} 
v_K(x,a,g_{s})=g_{s}^{-1}U_{K}(x,a) + U_{K}^{0}(x,a) + g_{s}U_{K}^{1}(x,a) +\dots 
\end{equation}   
is the~corresponding count of connected curves. In analogy to (\ref{eq:Psi_connected_expansion}), $U_K$ counts disks and $U_{K}^{k}$~counts curves of Euler characteristic $\chi = -k$.

\subsection{Curve counts at infinity}
\begin{figure}[htp]
	\centering
	%\includegraphics[width=.5\linewidth]{SFTHam}
	%\\
	\begin{tikzpicture}
		\draw (0, 0) -- (0, 4);
		\draw (1.5, 4) arc(0:-180:0.25 and 0.75);
		\draw (3, 4) arc(0:-180:0.25 and 0.75);
		\draw (2.5, 0) arc(0:180:0.5 and 1.5);
		\draw[green!80!black, -{Latex[length=1mm,width=2mm]}] (0.95, 4) -- (0.05, 4);
		\draw[green!80!black, -{Latex[length=1mm,width=2mm]}] (2.45, 4) -- (1.55, 4);
		\draw[green!80!black, -{Latex[length=1mm,width=2mm]}] (3.95, 4) -- (3.05, 4);
		\draw[green!80!black, -{Latex[length=1mm,width=2mm]}] (1.45, 0) -- (0.05, 0);
		\draw[green!80!black, -{Latex[length=1mm,width=2mm]}] (3.95, 0) -- (2.55, 0);

		\draw (3, 2.75) to [out=-45, in=180] (5, 2.75);
		\draw (5, 2.75) arc(90:-90:.5);
		\draw (3, 1.75) to [out=45, in=180] (5, 1.75);

		\draw (4.125, 2.30) arc(-125:-55:1);
		\draw (4.19, 2.255) arc(130:51:0.8);

		% \draw[red] (4.19, 0) -- (4.19, 4);
		% \draw[red] (3, 2.505) -- (6, 2.505);

		\node[green!80!black] at (0.5, 4.25) {$\beta$};
		\node[green!80!black] at (2, 4.25) {$\alpha_1$};
		\node[green!80!black] at (3.5, 4.25) {$\alpha_2$};
		\node[green!80!black] at (0.75, -0.25) {$\alpha_3$};
		\node[green!80!black] at (3.25, -0.25) {$\alpha_4$};
		\draw (4, 0) -- (4, 1.75);
		\draw (4, 2.75) -- (4, 4);

		\node[black] at (-1.5, 1) {$\Lambda_K \times \mathbb{R}$};
		\draw[-{Latex[length=1mm,width=1mm]}] (-1, 1.4) to [out=45,in=200] (-0.2, 2);
    \end{tikzpicture}
	\caption{A curve at infinity with positive degree one chord $\beta$, positive degree zero chords $\alpha_{1}$ and $\alpha_{2}$, and negative degree zero chords $\alpha_{3}$ and $\alpha_{4}$ contributes $g_{s}^{3}\alpha_{1}\alpha_{2}\partial_{\alpha_{3}}\partial_{\alpha_{4}}$ to $\mathbf{H}$.}
	\label{fig:SFTHam}
\end{figure}
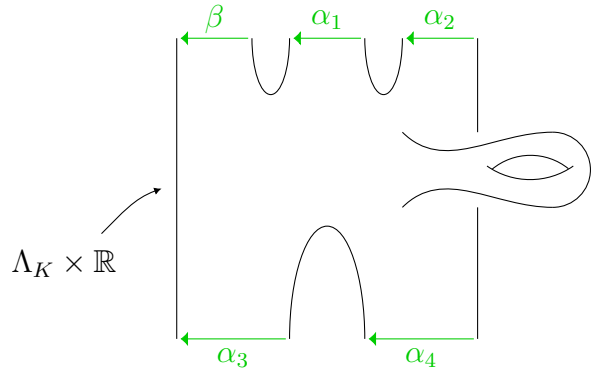

As explained in \cite{EN}, there is a~similar count of holomorphic curves at infinity, with boundary on $\Lambda_{K}\times\R$. 
Consider such curve with one positive degree one chord and the~rest degree zero chords. Recording degree zero punctures at positive infinity by variables $\alpha_{j}$ and negative infinity by dual differential operators $g_{s}\partial_{\alpha_{j}}$ (see Figure~\ref{fig:SFTHam}), we count curves at infinity, which leads to an~operator $\mathbf{H}$ on the~D-module with generators $\alpha_{j}, \partial_{\alpha_{j}}, \hat x, \hat y$. 
We write $f_{K}=\sum_{I} f_{K,I}\alpha_{I}$ for the~count of curves with positive punctures in the~monomials $\alpha_{I}=\alpha_{i_{1}}\dots \alpha_{i_{k}}$. Then our old potential is $f_{K,0}$ (i.e., $f_{K,0}=p_K$ for $L_K$ and $f_{K,0}=v_K$ for $M_K$) and we have
\begin{equation}
e^{-f_{K}}\, \mathbf{H} \,e^{f_{K}}=0,
\end{equation} 
since the~left hand side counts ends of a~1-dimensional moduli space. Eliminating $\alpha_{j}, \partial_{\alpha_{j}}$ from this equation, we find 
\begin{equation} 
\hat{A}_{K}(\hat x,\hat y,a) \;e^{f_{K,0}}=0.
\end{equation} 

\subsection{Annulus counts and the~Alexander polynomial}
In \cite{DE}, counts of holomorphic annuli stretching from $M_{K}$ to $S^{3}$ are considered and it is shown that the~Alexander polynomial and $A$-polynomial are related in the~following way:\footnote{Strictly speaking, here $A_{K}$ denotes the~augmentation polynomial, which is however conjectured to coincide with the~$A$-polynomial. For details and checks of the~conjecture see \cite{AV,FGS,FGSS,AENV,GKS,KS2}.}
\begin{equation}\label{eq:EkDi}
\Delta_{K}(x)=(1-x)\exp\left(\,\,\int\, \left.\frac{\partial_{\log a} A_{K}}{\partial_{\log y} A_{K}}\right|_{y=1,a=1}\,d\log x\right).
\end{equation}
Turning on $a$ and using the~parameterisation $y=e^{-\frac{\partial U_{K}}{\partial \log x}}$, we find that the~right hand side is the~$a\to 1$ limit of
\[ 
(1-x)\exp\left(-\frac{\partial U_{K}}{\partial \log a}(x)\right).
\] 
We will now consider the~counterpart of this equation for all orders in $g_{s}$. Fix $a=a_{0}$ and consider the~unnormalised expectation value of the~operator $e^{Ng_{s}\partial_{\log a}}$:
\begin{equation} 
\Theta_{N}(x,a_{0},q)=e^{Ng_{s}\partial_{\log a}}\exp\left(v_{K}(x,a_{0},g_s)\right)=\exp\left(v_{K}(x,q^Na_{0},g_s)\right).
\end{equation}
It then follows that $\Theta_{N}(x,a_{0},q)$ satisfies the~recursion:
\begin{equation} 
\hat{A}_{K}(\hat x,\hat y,q^Na_{0},q)\Theta_{N}(x,a_{0},q)=0.
\end{equation}
Then taking $a_{0}\to 1$ we get
\begin{equation}
\hat{A}_{K}(\hat x,\hat y,q^N,q)\Theta_{N}(x,1,q)=0.
\end{equation}
In other words, $\Theta_{N}(x,1,q)$ satisfies the~$SU(N)$-coloured HOMFLY-PT recursion and (after adjusting the~normalisation) 
we can identify it with the~$F_K$ invariant 
%\pk{Is it right to understand using N-1 instead of N in formulas above as considering SU(N) instead of U(N)?}:
\begin{equation}
\frac{\Delta_K(x)}{(1-x)^{N}}\cdot \Theta_{N}(x,1,q)
%{\nicefrac{\displaystyle\Theta_{1}(x,1,q)}{\displaystyle (1-x)}}=F_K^{SU(N),sym}(x,q).
%\frac{\Theta_{N}(x,1,q)}{\Theta_{1}(x,1,q)(1-x)^{N-1}}
=F_K^{SU(N),sym}(x,q).
\end{equation}

Finally, consider the~classical limit, corresponding to $g_s\to0$:
\begin{align}
\begin{split}\
&\langle e^{Ng_{s}\partial_{\log a}}\rangle = e^{-v}e^{Ng_{s}\partial_{\log a}} e^{v}, \\
&e^{-v} \left(1+N\frac{\partial U_{K}}{\partial \log a}+\frac12\left(N\frac{\partial U_{K}}{\partial \log a}\right)^{2}+\dots\right)e^{v} +O(g_{s}) =
e^{N\frac{\partial U_{K}}{\partial \log a}} +O(g_{s}).
\end{split}
\end{align}
Taking the~$a_{0}\to 1$ limit, we get 
\[ 
\left(\frac{1-x}{\Delta_{K}(x)}\right)^{N},
\]
and consequently
\[
\left(\frac{1}{\Delta_{K}(x)}\right)^{N-1}
\]
for the~normalised version in agreement with $\lim_{q\to 1}F_K^{SU(N),sym}(x,q)$ given by (\ref{eq:mixedlimit}).

%{\te{There is some strange normalization here that allows us to take the~limit $q\to 1$ and $a_0\to 1$ in such a~way that the~semi-classics matches on the~nose. Perhaps this is related to $SU(N)$ vs $U(N)$ but I am not sure.}}

\subsection{Geometric definitions of disk potentials and wave functions}

In this section we discuss geometric properties of the~curve counts involved in defining the~wave function (solutions to the~operator equation $\hat{A}_{K}\Phi_{K}(x)=0$) and their semiclassical analogues (solutions to the~equation $A_{K}=0$ of the~form $\log y=-\frac{\partial U_{K}}{\partial \log x}$, where $U_{K}=\sum_{r} c_{r}x^{r}$). 

In general one expects the~wave function to be a~count of all (generalised) disconnected holomorphic curves with boundary in $M_{K}$. Moreover, the~count of disconnected curves is given by the~exponential of the~count of connected curves, see equations (\ref{eq:Phi_full_expansion}-\ref{eq:Phi_connected_expansion}). 
%In other words
%\begin{equation}
%\Phi_{K} = \exp\left(g_{s}^{-1}U_{K} + U_{K}^{0} + g_{s}U_{K}^{1} +\dots + g_{s}^{k}U_{K}^{k} +\dots \right),
%\end{equation}  
%where $g_{s}$ is the~string coupling constant, $U_{K}$ counts disks, and $U_{K}^{k}$ counts curves of Euler characteristic $\chi = -k$. 
%As discussed above, we would like to count curves with boundary in $M_{K}$.
When $K$~is fibered, there exists a~non-vanishing $1$-form on $M_{K}$. We use this 1-form to shift $M_{K}\subset T^\ast S^3$  off of the~zero section $S^3$ and we consider the~curve counts above either in a~sufficiently SFT-stretched almost complex structure on $T^{\ast}S^{3}$ or in the~resolved conifold.

When $K$ is not fibered, there is no closed $1$-form on $M_{K}$ without zeros. It is straightforward to arrange that all zeros are critical points of index $1$ or $2$ (i.e. there are no local maxima or minima). In this case the~appropriate form of SFT-stretching leaves the~cotangent fiber in $T^{\ast}S^{3}$ at each critical point. When applying SFT-stretching, curves on the~outside may have punctures that end at Reeb chords stretching between fibers. Something similar happens with closed geodesics when stretching around manifolds other than $S^{3}$, see \cite{ES}. We describe how these curves could be taken into account. We consider first the~simpler case of the~disk potential, disregarding higher genus curves.

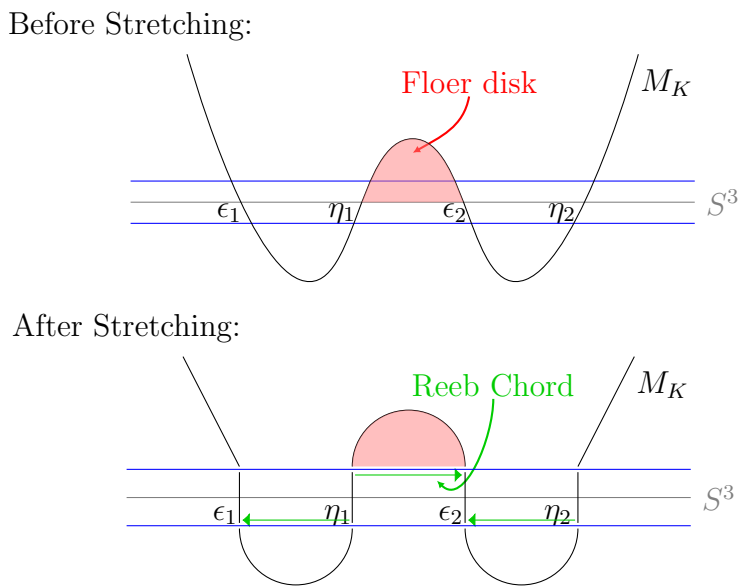
\begin{figure}[htp]
	\centering
	\begin{tikzpicture}[scale=0.75]
    	\node at (0, 3.5) {Before Stretching:};
    	\node at (9.5, 2.5) {$M_K$};
    	\node at (1.75, 3/16) {$\epsilon_1$};
    	\node at (3.75, 3/16) {$\eta_1$};
    	\node at (5.75, 3/16) {$\epsilon_2$};
    	\node at (7.65, 3/16) {$\eta_2$};
    	\node[gray] at (10.5, 0.375) {$S^3$};
    	\node[red] at (6, 2.5) {Floer disk};
    	\draw[thick, red, -{Latex[length=1mm,width=1mm]}] (6, 2.25) to [out=-100,in=30] (5, 1.25);
    	\draw[blue] (0,0) -- (10,0);
    	\draw[gray] (0,0.375) -- (10,0.375);
    	\draw[blue] (0,0.75) -- (10,0.75);
    	\draw[gray,name path=one, opacity=0] (4,0.375) -- (6,0.375);
    	\draw[name path=two] plot [smooth, tension=0.9] coordinates {(1,3) (3, -1) (5, 1.5) (7,-1) (9, 3)};
    	\tikzfillbetween[of=one and two,split,every even segment/.style = {opacity=0},
        every odd segment/.style  = {red!50, opacity=0.5}] {pattern=grid};
    \end{tikzpicture}
    \begin{tikzpicture}[scale=0.75]
    	\node at (0, 3.5) {After Stretching:};
    	\node at (9.5, 2.5) {$M_K$};
    	\node at (1.75, 3/16) {$\epsilon_1$};
    	\node at (3.75, 3/16) {$\eta_1$};
    	\node at (5.75, 3/16) {$\epsilon_2$};
    	\node at (7.65, 3/16) {$\eta_2$};
    	\node[green!80!black] at (6.5, 2.5) {Reeb Chord};
    	\draw[thick, green!80!black, -{Latex[length=1mm,width=1mm]}] (6.5, 2.25) to [out=-90,in=-45] (5.5, 0.8);
    	\draw[green!80!black, -{Latex[length=1mm,width=2mm]}] (3.95, 0.1) -- (2.05, 0.1);
    	\draw[green!80!black, -{Latex[length=1mm,width=2mm]}] (4.05, 0.9) -- (5.95, 0.9);
    	\draw[green!80!black, -{Latex[length=1mm,width=2mm]}] (7.95, 0.1) -- (6.05, 0.1);
    	\node[gray] at (10.5, 0.5) {$S^3$};
    	\draw[blue] (0,0) -- (10,0);
    	\draw[gray] (0,0.5) -- (10,0.5);
    	\draw[blue] (0,1) -- (10,1);
    	\draw (1, 3) -- (2, 1.05);
    	\draw (2, 0.05) -- (2, 0.95);
    	\draw (4, 0.05) -- (4, 0.95);
    	\draw (6, 0.05) -- (6, 0.95);
    	\draw (8, 0.05) -- (8, 0.95);
    	\draw (8, 1.05) -- (9, 3);
    	\draw (4, -0.05) arc(0:-180:1);
    	\draw[name path=two] (6, 1.05) arc(0:180:1);
    	\draw (8, -0.05) arc(0:-180:1);
    	\draw[blue,name path=one, opacity=0] (4,1.05) -- (6,1.05);
    	\tikzfillbetween[of=one and two,split,every even segment/.style = {red!50, opacity=0.5},
        every odd segment/.style  = {red!50, opacity=0.5}] {pattern=grid};
    \end{tikzpicture}
    \caption{The effect of SFT-stretching in the~presence of intersections with the~zero-section.}
	\label{fig:stretch}
\end{figure}

Write $\xi_{1},\dots,\xi_{r}$ and $\eta_{1},\dots,\eta_{r}$ for the~index $1$ and $2$ critical points in $M_{K}$, respectively. We write $\xi_{j}$ and $\eta_{j}$ also for the~corresponding cotangent fibers after stretching and $\partial\xi_{j}$ and $\partial \eta_{j}$ for their Legendrian boundary spheres, see Figure \ref{fig:stretch}. 
Fixing capping paths in the~Lagrangian, there is a~natural grading on the~Reeb chords which equals the~negative of the~dimension of a~disk with boundary in $M_{K}$ and negative puncture at the~Reeb chord. Then the~gradings are as follows:
\[ 
\begin{cases}
\partial\eta_{j} \to \partial\xi_{k}&\quad\text{ chord,}\quad \mathrm{grading} = 0,\\
\partial\eta_{j} \to \partial\eta_{k}&\quad\text{ chord,}\quad \mathrm{grading} = 1,\\
\partial\xi_{j} \to \partial\xi_{k}&\quad\text{ chord,}\quad \mathrm{grading} = 1,\\
\partial\xi_{j} \to \partial\eta_{k}&\quad\text{ chord,}\quad \mathrm{grading} = 2.
\end{cases}
\]

In the~disk potential $U_{K}$ we would now like to count not only actual closed disks but also disks with negative punctures at the~degree $0$ chords. Let $\alpha_{ij}$ denote the~degree $0$ chords, $\beta_{ij}$ and $\gamma_{ij}$ the~degree $1$ chords, and $\epsilon_{ij}$ the~degree $2$ chords. Here $\beta_{ii}$ and $\gamma_{ii}$ are formal length zero chords associated with boundaries of bounding chains. We next note that such counts cannot be invariant under deformations for the~following reason.

\begin{figure}[htp]
	\centering
	\begin{tikzpicture}[scale=0.75]
    	\begin{scope}[decoration={
    	    markings,
    	    mark=at position 0.5 with {\arrow{latex}}}
    	    ]
    		
    		\draw[green!80!black, postaction={decorate}] (4 + 4, 0) -- (4 + 0, 0);
    		\draw[red] (4 + 4, 0) -- (4 + 4, 4);
    		\draw (4 + 4, 4.05) arc(0:180:0.875 and 1.25);
    		\draw[green!80!black, postaction={decorate}] (4 + 4, 4) -- (4 + 2.25, 4);
    		\draw[red] (4 + 2.25, 4) arc(0:-180:0.25 and 1.5);
    		\draw[green!80!black, postaction={decorate}] (4 + 1.75, 4) -- (4 + 0, 4);
    		\draw (4 + 0, 4.05) arc(180:0:0.875 and 1.75);
    		\draw[blue] (4 + 0, 4) -- (4 + 0, 0);
    
    		\node at (4 + 2, 5.5) {$M_K$};
    		\node at (4 + 4.25, 5.25) {$M_K$};
    		\node at (4 + -2, 5) {dim $= 0$};
    		\node at (4 + 6.5, 5) {dim $= -1$};
    		\node at (4 + -2, 1) {dim $= 1$};
    		\draw[-{Latex[length=1mm,width=1mm]}] (4.2 + -1, 4.9) to [out=-30, in=-150] (4 + 0.5, 5);
    		\draw[-{Latex[length=1mm,width=1mm]}] (3.8 + 5.25, 4.9) to [out=-150, in=-30] (4 + 3.5, 4.5);
    		\draw[-{Latex[length=1mm,width=1mm]}] (4.2 + -1, 0.9) to [out=0, in=-135] (4 + 0.5, 2);
    
    		\node[red] at (4 + 2, 2.125) {$\eta$};
    		\node[red] at (4 + 4.25, 2) {$\eta$};
    		\node[blue] at (4 + -0.25, 2) {$\xi$};
    		\node[green!80!black] at (4 + 1, 3.65) {$\alpha$};
    		\node[green!80!black] at (4 + 2, 0.25) {$\alpha$};
    		\node[green!80!black] at (4 + 3, 3.65) {$\beta$};
    
    		\node[yellow!50!orange, align=left] at (2, -2) {Contribution to \\ differential};
    		\node[yellow!50!orange, align=left] at (10, -1.75) {Change in \\ potential};
    
    		\draw[thick, yellow!50!orange, -{Latex[length=1mm,width=1mm]}] (5, -0.5) to [out=-90, in=90] (3, -5);
    
    		\draw[thick, yellow!50!orange, -{Latex[length=1mm,width=1mm]}] (7, -0.5) to [out=-90, in=90] (10, -4.5);
    
    		\draw[green!80!black, postaction={decorate}] (4, -8 - 2) -- (0, -8 - 2);
    		\draw[gray] (4, -8 - 2) -- (4, -4 - 2);
    		\draw[green!80!black, postaction={decorate}] (4, -4 - 2) -- (2.25, -4 - 2);
    		\draw[gray] (2.25, -4 - 2) arc(0:-180:0.25 and 1.5);
    		\draw[green!80!black, postaction={decorate}] (1.75, -4 - 2) -- (0, -4 - 2);
    		\draw (0, -3.95 - 2) arc(180:0:0.875 and 1.75);
    		\draw[gray] (0, -4 - 2) -- (0, -8 - 2);
    
    		\node at (2, 5.5 - 10) {$M_K$};
    		\node at (-2, 5 - 10) {dim $= 0$};
    		\node at (-2, 1 - 10) {dim $= 1$};
    		\draw[-{Latex[length=1mm,width=1mm]}] (-0.8, 4.9 - 10) to [out=-30, in=-150] (0.5, 5 - 10);
    		\draw[-{Latex[length=1mm,width=1mm]}] (-0.8, 0.9 - 10) to [out=0, in=-135] (0.5, 2 - 10);
    
    		\node[gray] at (2, 2.125 - 10) {$\eta$};
    		\node[gray] at (4.25, 2 - 10) {$\eta$};
    		\node[gray] at (-0.25, 2 - 10) {$\xi$};
    		\node[green!80!black] at (1, 3.65 - 10) {$\alpha$};
    		\node[green!80!black] at (2, 0.25 - 10) {$\alpha$};
    		\node[green!80!black] at (3, 3.65 - 10) {$\beta$};
    
    		\draw[green!80!black, postaction={decorate}] (12, -8 - 2) -- (8, -8 - 2);
    		\draw (12, -8 - 2) -- (12, -4 - 2);
    		\draw (12, -4 - 2) arc(0:180:0.875 and 1.25);
    		\draw (10.25, -4 - 2) arc(0:-180:0.25 and 1);
    		\draw (8, -4 - 2) arc(180:0:0.875 and 1.75);
    		\draw (8, -4 - 2) -- (8, -8 - 2);
    		
    		\node at (12, -2.5 - 2) {$M_K$};
    		\node[green!80!black] at (10, -7.75 - 2) {$\alpha$};
    	\end{scope}
    \end{tikzpicture}
	\caption{Counts of disks with negative punctures changes at instances where there are $(-1)$-disks. In order to get an~invariant, we use augmentations of a~differential graded algebra with differential that counts the~$(+1)$-part attached to the~$(-1)$-disk.}
	\label{fig:differential}
\end{figure}
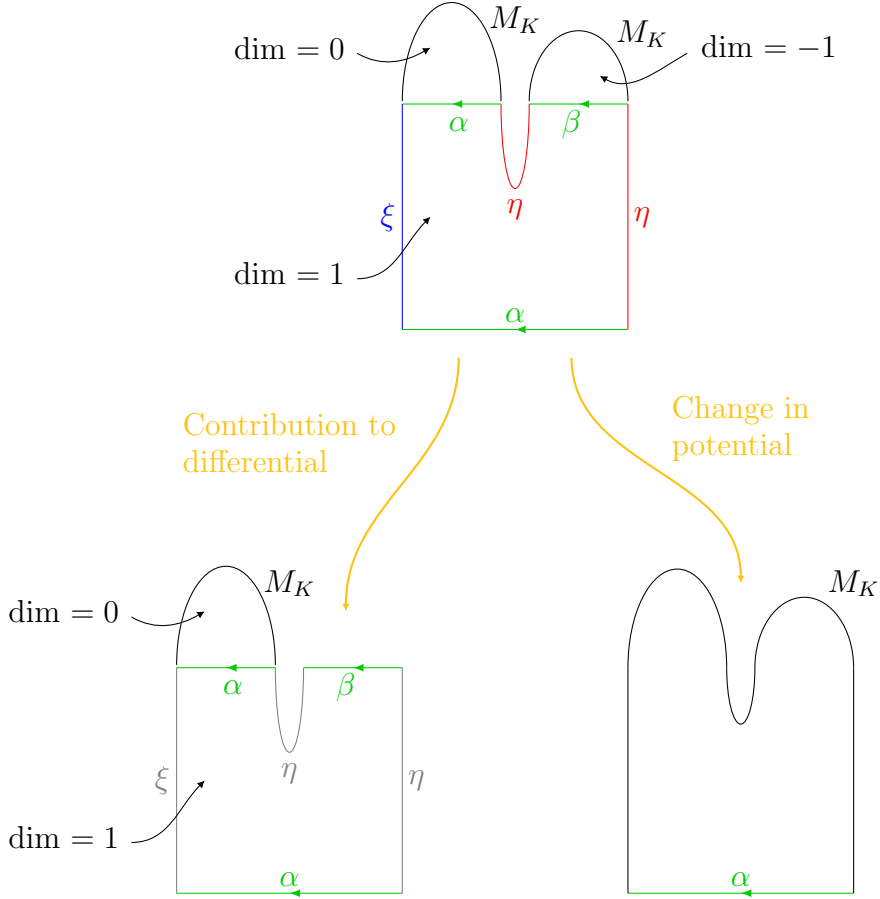

In a~generic 1-parameter family there might appear isolated instances where there is a~disk~$\sigma$ of dimension $-1$. Such disk has negative punctures at one $\beta_{ij}$ or $\gamma_{ij}$ chord and the~rest at $\alpha_{ij}$-chords. At such an~instance, the~count of disks changes by gluing to $\sigma$ a~disk with positive puncture at the~degree~$1$ chord and positive and negative punctures at degree~0 chords, where all positive punctures are capped off by disks with corresponding negative punctures, see Figure \ref{fig:differential}.
This indicates that one should count the~disks not with coefficients in the~$\alpha_{ij}$, but rather in an~augmentation of the~differential graded algebra with differential given by disks with positive degree 0 punctures capped off, see Figure \ref{fig:differential}.  Note that this algebra in degree $0$ has linearised homology which is a~torsion $x^{\pm 1}$-module. Its augmentation variety therefore defines $\alpha_{ij}$ as a~function of $a$ and~$x$. This variety may have many branches and correspondingly we get several disk potentials.

This raises the~question which branch is the~right one to give the~Alexander polynomial, according to the~formula \eqref{eq:EkDi}. Although it is not easy to characterise that branch concretely, %section 6 in 
\cite{DE} shows that such a~branch exists as follows. The~Alexander polynomial is given by a~product
\begin{equation}
\Delta_{K}(x)=(1-x)\exp(B_{K}(x))\det(D_{K}(x)),
\end{equation}
where $D_{K}(x)=D_{K,0}+O(x)$ is the~differential on the~Morse-Novikov complex of $M_{K}$ viewed as an~$\C[x^{\pm 1}]$-module, and $B_K(x)$ is the~count of holomorphic annuli stretching between $M_K$ and $S^3$. We point out that the~coefficient $\Delta_{K,0}$ of the~leading term in the~Alexander polynomial is 
\[
\Delta_{K,0}=\det(D_{K,0}),
\]
which is equal to $1$ for fibered knots, see the~discussion in Section \ref{ssec:5_2}.
%\pk{Comparing with (35) I think it should be $U^0_K(x)$ instead of $V^0_K(x)$ -- earlier it was $B_K(x)$}
%\te{No, it is not the~same thing as $U^0_K$ that counts annuli with boundary on $M_K$ only (and possibly punctures) after stretching.}
Here the~left hand side remains constant under deformations. %In the~left hand side, 
On the~right hand side, at non-generic instances factors may move from the~second to the~third factor or in the~opposite direction. For sufficiently stretched almost complex structures there can be no further moving of factors and therefore $\log\det(D_{K}(x))$ should be the~contribution to $\frac{\partial U_{K}}{\partial \log a}$ coming from disks with additional punctures.  

As an~illustration of the~differential graded algebra at the~negative end, we consider the~basic case when we add two canceling critical points of the~shifting $1$-form. This leads to an~algebra with one chord $\alpha$ of degree $0$, two chords $\beta$ and $\gamma$ of degree $1$, and one chord $\epsilon$ of degree~$2$. The~relevant part of the~differential is related to the~Floer disk that cancels the~two critical points. After stretching, this Floer disk gives a~disk with a~negative puncture at~$\alpha$ and homologically trivial boundary. In analogy with ordinary disks, one expects that all its multiple covers contribute to the~differential, and taking the~puncture into account one gets the~count
\[
\sum_{k=1}^{\infty} \;\alpha^{k} \ =  \ \frac{\alpha}{1-\alpha}.
\]
An augmentation must vanish on the~image of the~differential and the~augmentation variety is then given by the~equation
\begin{equation}
0=\frac{\alpha}{1-\alpha}\quad\text{ or }\quad\alpha=0.
\end{equation}
It follows that the~disk potential remains unchanged, as expected.

In the~higher genus case, one should upgrade the~differential graded algebra at the~negative end just described to an~SFT structure. More precisely, capping off with all genus curves instead of only disks, one finds an~operator $\mathbf{H}$ that counts curves with one positive puncture at $\beta_{ij}$-chord and other punctures at $\alpha_{ij}$-chords. We require that 
\begin{equation}
e^{-f_{\alpha}}\,\mathbf{H}\, e^{f_{\alpha}} = 0.
\end{equation} 
Eliminating $\alpha_{ij}$ and $\partial_{\alpha_{ij}}$, we get $\alpha_{ij}=\alpha_{ij}(y,q)$ and possibly non-unique wave functions corresponding to different solutions.

\section{An \texorpdfstring{$a$}{a}-deformation of \texorpdfstring{$F_K$}{FK}}\label{sec:a-deformed F_K}
In this section we present our main results on $a$-deformed $F_{K}$ invariants -- analogues of HOMFLY-PT polynomials for 3-manifolds with the~topology of the~knot  complement. For simplicity we will usually refer only to the~knot $K$, having in mind that the~corresponding $3$-manifold is $M_K=S^{3}\backslash K$. 

A few remarks on the~convention we use are in order:
\begin{itemize}
    \item In the~literature one can find three different normalisations corresponding to different values of $F_K$ for the~unknot:
    \begin{itemize}
        \item In {\it reduced} normalisation it is simply set to be $1$. This normalisation is present in the~majority of papers on HOMFLY-PT, superpolynomials, and $A$-polynomials, e.g. \cite{DGR,FGSA,FGS,FGSS,NRZS}.
        \item In {\it unreduced} normalisation it is equal to \eqref{F_0_1} -- the~numerator of the~full unknot factor. This convention is dominant in the~growing literature on $F_K$ invariants, e.g.~\cite{GM,Park1,Park2,GHNPPS}, usually combined with the~balanced expansion (see the~next bullet point).
        \item In {\it fully unreduced} normalisation it is equal to the~full unknot factor \eqref{eq: F_0_1 fully unred}. This normalisation is natural in the~context of enumerative invariants and can be found in \cite{OV,AENV,EN,EKL1,EKL2,ES,DE}. In the~literature this normalisation is usually called just ``unreduced", but since we join different perspectives, we have to distinguish it from the~one discussed in the~previous point.  
    \end{itemize}
     We present our results mostly in the~reduced normalisation. In case of the~geometric considerations in Sections \ref{sec:Enumerative geometry} and \ref{sec:Quantum modularity} we analyse curve counts leading to fully unreduced normalisation and explain how to obtain the~reduced one. Conjecture \ref{$a$-deformed $F_K$} is formulated in the~reduced normalisation except for \eqref{eq:specializetoSUN} which should be compared in the~unreduced normalisation. 
    %\item There are different normalisations, depending on whether we set $F_K$ for the~unknot to be $1$ or not. We refer to the~former as \emph{reduced} normalisation and the~latter as \emph{unreduced} normalisation. In other words, the~unreduced version is simply the~reduced version times the~unknot factor \eqref{F_0_1}. We present our results mostly in the~reduced normalisation except for the~unknot. Conjecture \ref{$a$-deformed $F_K$} is formulated in the~reduced normalisation except for \eqref{eq:specializetoSUN} which should be compared in the~unreduced normalisation. 
    \item We use the~\emph{positive expansion} of $F_K$, meaning that we express $F_K$ as a~power series in $x$ expanded around $x=0$. To get the~\emph{negative expansion}, the~power series expanded around $x=\infty$ we can simply use the~Weyl symmetry \eqref{Weyl_symmetry}. The~\emph{balanced expansion} such as the~one in \eqref{eq:GMconj} is simply the~average of the~positive and negative expansions. 
   % \item For simplicity, we shift the~overall $x$-degree so that $F_K$ starts from a~constant term. To undo this shift, one can simply multiply the~shifted one by $x^{\frac{\log a}{\hbar}~-1}$. This shift is reflected in the~fact that we rescale $\hat{y}$ by $a/q$ in the~quantum $A$-polynomial. \pk{Works for the~trefoil, for general torus knots the~prefactor is $a^{pr} q^{-pr} = x^{p(\log a/\hbar-1)} $}
\end{itemize}
%In all cases we adapt the~convention in which we expand $F_{K}$ around $x=0$ (it means that we consider only the~\emph{positive expansion} -- the~one that corresponds to the~non-negative powers of $x$), as well as shift the~overall $x$-degree so that $F_K$ starts from a~constant term.\footnote{Of course, we can always recover the~usual \emph{balanced expansion} using the~Weyl symmetry \eqref{Weyl_symmetry}. For this, first multiply by $x^{\log a~-1}$ (since the~convention we are using in this section is different from the~one in Conjecture~\ref{$a$-deformed $F_K$}), and then apply the~Weyl symmetry \eqref{Weyl_symmetry}.} Such choice is very helpful in working with HOMFLY-PT and $A$-polynomials. In case of the~unknot we consider the~unreduced normalisation, for other knots we divide by the~unknot factor which corresponds to the~reduced normalisation.
In order to familiarise the~reader with all subtleties associated to conventions, in the~next section we present them in the~example of the~unknot.

\subsection{Unknot}\label{sec:a-deformed unknot results}

In case of the~unknot the~$a$-deformed $F_{K}$ invariant can be obtained from the~natural HOMFLY-PT polynomial in representation $S^{r}$. It is given by \cite{FGS}
\begin{equation}
\bar{P}_{r}(0_{1};a,q)=a^{-\frac{r}{2}}q^{\frac{r}{2}}\frac{(a;q)_{r}}{(q;q)_{r}}=a^{-\frac{r}{2}}q^{\frac{r}{2}}\frac{(a;q)_{\infty}(q^{r+1};q)_{\infty}}{(aq^{r};q)_{\infty}(q;q)_{\infty}},
\end{equation}
where
\begin{equation}
(z;q)_{n}=\prod_{i=1}^{n-1}(1-zq^{i})
\end{equation}
is the~$q$-Pochhammer symbol. The~expression \eqref{eq:reduced vs unreduced} corresponds to $r=1$.

After performing the~substitution $q^{r}=x$, we obtain the~$F_{K}$ invariant in the~fully unreduced normalisation:
\begin{equation}\label{eq: F_0_1 fully unred}
F^{\textrm{full.unred.}}_{0_{1}}(x,a,q)=a^{-\frac{\log x}{2\hbar}}x^{\frac{1}{2}}\frac{(a;q)_{\infty}(xq;q)_{\infty}}{(xa;q)_{\infty}(q;q)_{\infty}}=x^{\frac{-\log a}{2\hbar}+\frac{1}{2}}\frac{(xq;q)_{\frac{\log a}{\hbar}-1}}{(q;q)_{\frac{\log a}{\hbar}-1}}.
\end{equation}
For $a=q^N$ it reduces to
\begin{equation}
F^{\textrm{full.unred.}}_{0_{1}}(x,q^N,q)=x^{-\frac{N-1}{2}}\frac{(xq;q)_{N-1}}{(q;q)_{N-1}},
\end{equation}
which for $SU(2)$ gives
\begin{equation}\label{eq:SU(2) unknot fully unreduced}
F^{\textrm{full.unred.}}_{0_{1}}(x,q^2,q)=\frac{(xq)^{1/2}-(xq)^{-1/2}}{q^{1/2}-q^{-1/2}}.
\end{equation}

We obtain the~$F_{K}$ invariant in the~unreduced normalisation by dropping the~prefactors:
\begin{equation}\label{F_0_1}
F^{\textrm{unred.}}_{0_{1}}(x,a,q)=\frac{x^{\frac{-\log a}{2\hbar}+\frac{1}{2}}\frac{(a;q)_{\infty}(xq;q)_{\infty}}{(xa;q)_{\infty}(q;q)_{\infty}}}{x^{\frac{-\log a}{2\hbar}+\frac{1}{2}}\frac{(a;q)_{\infty}}{(q;q)_{\infty}}}=\frac{(xq;q)_{\infty}}{(xa;q)_{\infty}}=(xq;q)_{\frac{\log a}{\hbar}-1}.
\end{equation}
Substituting $a=q^N$ and $a=q^2$, we get
\begin{equation}
F^{\textrm{unred.}}_{0_{1}}(x,q^N,q)=(xq;q)_{N-1},    \qquad \qquad F^{\textrm{unred.}}_{0_{1}}(x,q^2,q)=1-xq,
\end{equation}
so we can see that this unknot factor is appropriate for the~positive expansion that we use in the~paper.
The~balanced expansion requires keeping the~prefactor $-x^{\frac{-\log a}{2\hbar}+\frac{1}{2}}$, which for $SU(2)$ leads to symmetric expression $(xq)^{1/2}-(xq)^{-1/2}$, being the~numerator of \eqref{eq:SU(2) unknot fully unreduced}. Taking into account the~fact that here $x=q^r$ for $S^r$, whereas in \cite{GM} $x=q^n$ for $S^{n-1}$, one can recognise the~familiar factor $x^{1/2}-x^{-1/2}$ corresponding to switching between reduced and unreduced $F^{SU(2)}_K$.

Finally, the~reduced $F_{K}$ invariant is simply set to 1. Following \eqref{eq:reduced vs unreduced}, one can also say that it is obtained by the~division by the~full unknot factor:
\begin{equation}
    F^{\textrm{red.}}_{0_{1}}(x,a,q)=\frac{x^{\frac{-\log a}{2\hbar}+\frac{1}{2}}\frac{(a;q)_{\infty}(xq;q)_{\infty}}{(xa;q)_{\infty}(q;q)_{\infty}}}{F^{\textrm{full.unred.}}_{0_{1}}(x,a,q)}=1.
\end{equation}

Since the~unreduced normalisation is absent in the~literature on the~$A$-polynomials, let us analyse the~recursion $\hat{A}^{\textrm{unred.}}_{0_{1}}(\hat{x},\hat{y},a,q)F^{\textrm{unred.}}_{0_{1}}(x,a,q)=0$. We have
\begin{equation}\label{eq:Ahat 0_1}
    \hat{A}^{\textrm{unred.}}_{0_{1}}(\hat{x},\hat{y},a,q)=(1-a\hat{x})-(1-q\hat{x})\hat{y},
\end{equation}
which agrees with the~quantum $a$-deformed $A$-polynomial from \cite{FGS} after taking into account dropping the~prefactor and the~conventional difference $\hat{x}_{\textrm{FGS}}=\hat{x}q$.

%Since $F_{0_{1}}(x,a,q)=\left.P_{r}(0_{1};a,q)\right|_{q^{r}=x}$, its 
On the~other hand, the~semiclassical limit of $F_K$ reproduces the~twisted superpotential of the~3d $\mathcal{N}=2$ theory associated to the~unknot complement and analysed in~\cite{FGS}:
\begin{equation}
\begin{split} & F^{\textrm{unred.}}_{0_{1}}(x,a,q)\underset{\hbar\rightarrow0}{\rightarrow}\exp\left[\frac{1}{\hbar}\widetilde{\mathcal{W}}(x,a)+O(\hbar^{0})\right],\\
 & \widetilde{\mathcal{W}}(x,a)=\textrm{Li}_{2}(x)-\textrm{Li}_{2}(ax).
\end{split}
\label{eq:Semiclassical limit F_unknot}
\end{equation}
Introducing the~variable $y$ dual to $x$ we obtain
\begin{equation}
\log y=\frac{\partial\widetilde{\mathcal{W}}(x,a)}{\partial\log x}=\log\left(1-ax\right)-\log\left(1-x\right).
\end{equation}
In compliance with Section \ref{sec:Physical}, this equation is equivalent to the~zero locus of the~$A$-polynomial:
\begin{equation}
A^{\textrm{unred.}}_{0_{1}}(x,y,a)=1-ax-y+xy,
\end{equation}
which is a~classical limit of (\ref{eq:Ahat 0_1}).

\subsection{Trefoil knot}
Similarly to the~unknot case, we can obtain the~closed form expression for the~$a$-deformed $F_{K}$ invariant for the~trefoil using the~formula for (reduced) HOMFLY-PT polynomial with $q^{r}=x$. Starting from \cite{FGS}
\begin{equation}
P_{r}(3_{1};a,q)=\sum_{k=0}^{r}a^{r}q^{r(k-1)+k}\frac{(q^{r};q^{-1})_{k}(aq^{-1};q)_{k}}{(q;q)_{k}},
\end{equation}
changing the~summation to the~infinity ($(q^{r};q^{-1})_{k}$ vanishes for $k > r$), and substituting $q^{r}$ by~$x$, we obtain
\begin{equation}\label{eq:Trefoil with prefactor}
   \sum_{k=0}^{\infty}x^{\frac{\log a}{\hbar}-1}(xq)^{k}\frac{(x;q^{-1})_{k}(aq^{-1};q)_{k}}{(q;q)_{k}}. 
\end{equation}
For simplicity, we omit the~prefactor $x^{\frac{\log a}{\hbar}-1}$:
\begin{equation}\label{eq:F_3_1}
F_{3_{1}}(x,a,q)=\sum_{k=0}^{\infty}(xq)^{k}\frac{(x;q^{-1})_{k}(aq^{-1};q)_{k}}{(q;q)_{k}},
\end{equation}
however it is important to keep in mind that we need it in order to use the~Weyl symmetry~\eqref{Weyl_symmetry}.
In the~known results for particular $N$, such as \cite{GM,Park1}, this prefactor reduces to $x^{N-1}$ and is kept explicit.

The function $F_{3_{1}}(x,a,q)$ is annihilated by the~quantum $a$-deformed A-polynomial
\begin{equation}
\label{eq:A-polynomial for trefoil}
\hat{A}_{3_{1}}(\hat{x},\hat{y},a,q)=a_{0}+a_{1}\hat{y}+a_{2}\hat{y}^{2},
\end{equation}
where\footnote{Formula (\ref{eq:A-polynomial for trefoil}) differs from \cite{FGS} by the
rescaling of $\hat{x}$ by $q$ mentioned earlier, and of $\hat{y}$ by $a/q$ due to the~omitted prefactor.}
\begin{equation*}
\begin{split}
a_{0}= & q^4\hat{x}^{3}(q\hat{x}-1)(1-aq^3\hat{x}^{2}),\\
a_{1}= & -(1-aq^2\hat{x}^{2})(1+q^{4}\hat{x}^{2}-aq\hat{x}^{2}+a^{2}q^4\hat{x}^{4}+q^{2}\hat{x}(-1+q\hat{x}-aq\hat{x}-aq^2\hat{x}^{2})),\\
a_{2}= & (1-aq\hat{x})(1-aq\hat{x}^{2}).
\end{split}
\end{equation*}

Similarly to the~unknot case, the~semiclassical limit of  $F_{3_{1}}(x,a,q)$ reproduces the~twisted superpotential of $T[M_{3_{1}}]$, the~trefoil complement theory studied in \cite{FGS}:
\begin{equation}
\begin{split} & F_{3_{1}}(x,a,q)\underset{\hbar\rightarrow0}{\rightarrow}\exp\int\frac{dz}{z}\left[\frac{1}{\hbar}\widetilde{\mathcal{W}}(z,x,a)+O(\hbar^{0})\right],\\
 & \widetilde{\mathcal{W}}(z,x,a)=\log x\log z-\textrm{Li}_{2}(x)+\textrm{Li}_{2}(xz^{-1})+\textrm{Li}_{2}(a)-\textrm{Li}_{2}(az)+\textrm{Li}_{2}(z).
\end{split}
\label{eq:Semiclassical limit F_trefoil}
\end{equation}
Extremalisation with respect to $z$ ($z_{0}$ denotes the~extremal value) and introduction of the~variable $y$ dual to $x$ leads to equations
\begin{equation}
\begin{cases}
1=\frac{x\left(1-xz_{0}^{-1}\right)\left(1-az_{0}\right)}{\left(1-z_{0}\right)},\\
y=\frac{z_{0}(1-x)}{1-xz_{0}^{-1}}.
\end{cases}
\end{equation}
Eliminating $z_{0}$, we obtain the~$A$-polynomial
\begin{equation}
A_{3_{1}}(x,y,a)=(x-1)x^{3}-\left(1-x+2(1-a)x^{2}-ax^{3}+a^{2}x^{4}\right)y+(1-ax)y^{2},
\end{equation}
which is the~classical limit of (\ref{eq:A-polynomial for trefoil}).

We would like to compare our results with $F_{K}$ invariants for $SU(2)$ case from \cite{GM} and $SU(N)$ case from \cite{Park1}. For this, we need to prepare $F_K$ for the~right-handed trefoil (which is the~mirror of left-handed trefoil we presented above) in the~unreduced normalisation, rescaling $x$, keeping the~prefactor $x^{N-1}$, and using the~balanced expansion (see Section~\ref{sec:a-deformed unknot results} and remarks before it). Let us go through all this conventional changes step by step for the~$SU(2)$ case; for $SU(N)$ the~method is completely analogous.

We start from the~formula \eqref{eq:Trefoil with prefactor} which contains the~prefactor $x^{N-1}=x$ and set $a=q^2$ to obtain
\begin{equation} \label{eq:F_3_1^l for SU(2) with prefactor}
\sum_{k=0}^{\infty}x(xq)^{k}(x;q^{-1})_{k}.
\end{equation}
In order to get the~formula for the~right-hand trefoil, we have to perform the~change of variables
\[
x\mapsto x^{-1},\qquad q\mapsto q^{-1}%,\qquad a\mapsto a^{-1}
\]
and use the~Weyl symmetry
\[
x^{-1}\mapsto ax=q^2 x,
\]
which gives
%\begin{equation} \label{F_3_1^r}
%F_{3_{1}^{r}}(x,a,q)=\sum_{k=0}^{\infty}(xaq^{-1})^{k}\frac{(xa;q)_{k}(a^{-1}q;q^{-1})_{k}}{(q^{-1};q^{-1})_{k}}.
%\end{equation}
%In order to relate this to the~known results from \cite{GM} and \cite{Park1}, we still need to align the~normalisation and conventions (see Section \ref{sec:a-deformed unknot results} and remarks before it). Here we briefly describe the~required method %for the~$SU(2)$ case -- for $SU(N)$ all steps are completely analogous. %It is similar to compare them in the~$SU(N)$ case, with unreduced normalisation.
%We start from setting $a \mapsto q^2$, corresponding to $SU(2)$, and restoring the~prefactor $x^{\frac{\log a}{\hbar}-1}=x$, which is transformed to $x\mapsto x^{-1}\mapsto ax=q^2 x$ due to considering the~mirror image. This changes \eqref{F_3_1^r} into
\begin{equation} \label{eq:F_3_1^r for SU(2) with prefactor}
\sum_{k=0}^{\infty}xq^2(xq)^{k}(xq^2;q)_{k}.
\end{equation}
Then we have to switch from $x=q^r$ for $S^r$ to $x=q^n$ for $S^{n-1}$ used in \cite{GM}, which corresponds to $x\mapsto x/q$. Performing this transformation and expanding in $x$ we get
% xq[   1 + x + (1 - q)x^2 + (1 - q - q^2)x^3 + (1 - q - q^2 + q^5)x^4 + \ldots]
\begin{equation}
qx + qx^2 + q(1 - q)x^3 + q(1 - q - q^2)x^4 + q(1 - q - q^2 + q^5)x^5 + \ldots.
\end{equation}
Since \cite{GM} uses the~unreduced normalisation, 
%In the~conventions of \cite{GM} what we have computed is a~reduced version of the~asymmetric expansion of $f_{3_1^r}^{SU(2)} = F_{3_1^r}^{SU(2)}/(x^{1/2} - x^{-1/2})$. 
we multiply by $(x^{\frac{1}{2}} - x^{-\frac{1}{2}})$, which leaves us with the~series
\begin{equation}
    -qx^{\frac{1}{2}} + q^2x^{\frac{5}{2}} + q^3x^{\frac{7}{2}} - q^6x^{\frac{11}{2}} - q^8x^{\frac{13}{2}} + \ldots    
\end{equation}
From here we simply need to switch to the~balanced expansion, which -- thanks to the~Weyl symmetry -- means replacing $x^n \mapsto \frac{1}{2}(x^{n} - x^{-n})$ for $SU(2)$, and we exactly recover ($q,x$)-series from  \cite[eq.(114)]{GM}:
\begin{equation}
\begin{split}
    F_{3_{1}^{r}}^{SU(2)}(x,q) =&
    -\frac{1}{2} \Big[q(x^{\frac{1}{2}} - x^{-\frac{1}{2}}) - q^2(x^{\frac{5}{2}} - x^{-\frac{5}{2}}) - q^3(x^{\frac{7}{2}} - x^{-\frac{7}{2}})\\& + q^6(x^{\frac{11}{2}} - x^{-\frac{11}{2}}) + q^8(x^{\frac{13}{2}} - x^{-\frac{13}{2}}) + \ldots \Big].
\end{split}
\end{equation}

Following the~same procedure, we find a~perfect agreement with the~results of \cite{Park1}, presented below for $N=3,4$ in the~unreduced normalisation and using balanced expansion. Here $\cong$ denotes equality up to sign and a~multiplication by a~monomial $q^d$ for some $d\in \mathbb{Q}$. 
\begin{align}
    F_{3_1^r}^{\mathrm{sym,unred}, SU(3)}(x,q) &\cong \frac{1}{2}\Big[(q^{1/2}x+q^{-1/2}x^{-1})(1)\nonumber\\ 
    &+ (q^{3/2}x^3+q^{-3/2}x^{-3})(-q-q^2)\nonumber\\ 
    &+ (q^2x^4+q^{-2}x^{-4})(-q^{5/2}-q^{7/2})\\
    &+ (q^{5/2}x^5+q^{-5/2}x^{-5})(q^3)\nonumber\\
    &+ (q^{3}x^6+q^{-3}x^{-6})(q^{9/2}+q^{11/2}+q^{13/2}+q^{15/2})+\cdots\Big],\nonumber\\
    F_{3_1^r}^{\mathrm{sym,unred}, SU(4)}(x,q) &\cong \frac{1}{2}\Big[(q^{3/2}x^{3/2}-q^{-3/2}x^{-3/2})(1)\nonumber\\
    &+ (q^{7/2}x^{7/2}-q^{-7/2}x^{-7/2})(-q-q^2-q^3)\nonumber\\
    &+ (q^{9/2}x^{9/2}-q^{-9/2}x^{-9/2})(-q^3-q^4-q^5)\\
    &+ (q^{11/2}x^{11/2}-q^{-11/2}x^{-11/2})(q^3+q^4+q^5)\nonumber\\
    &+ (q^{13/2}x^{13/2}-q^{-13/2}x^{-13/2})(q^5+2q^6+2q^7+2q^8+q^9+q^{10}) +\cdots\Big].\nonumber
\end{align}

\subsection{\texorpdfstring{$(2,2p+1)$}{(2,2p+1)} torus knots}\label{sec:Torus knots}

We can generalise the~results for the~trefoil to all $(2,2p+1)$ torus knots. Basing on \cite{FGSS}, we have
\begin{equation}
\begin{split}
P_{r}(T^{(2,2p+1)};a,q)=\sum_{0\leq k_{p}\leq\ldots\leq k_{1}\leq k_{0}=r} & a^{pr}q^{-pr} q^{(2r+1)(k_{1}+k_{2}+\ldots+k_{p})-\sum_{i=1}^{p}k_{i-1}k_{i}} \\
\times & \frac{(q^{r};q^{-1})_{k_{1}}(aq^{-1};q)_{k_{1}}}{(q;q)_{k_{1}}}\left[\begin{array}{c}
k_{1}\\
k_{2}
\end{array}\right]\cdots\left[\begin{array}{c}
k_{p-1}\\
k_{p}
\end{array}\right],
\end{split}
\end{equation}
where we use the~$q$-binomial
\begin{equation}
\left[\begin{array}{c}
n\\
k
\end{array}\right]=\frac{(q;q)_{n}}{(q;q)_{k}(q;q)_{n-k}}.
\end{equation}
Changing the~summation to infinity and substituting $q^{r}=x$ leads to
\begin{equation}\label{eq:F_K for torus knots with prefactor}
\begin{split}\sum_{0\leq k_{p}\leq\ldots\leq k_{1}} & x^{p(\frac{\log a}{\hbar}-1)} x^{2(k_{1}+\ldots+k_{p})-k_{1}}q^{(k_{1}+k_{2}+\ldots+k_{p})-\sum_{i=2}^{p}k_{i-1}k_{i}}\\
\times & \frac{(aq^{-1};q)_{k_{1}}(x;q^{-1})_{k_{1}}}{(q;q)_{k_{1}}}\left[\begin{array}{c}
k_{1}\\
k_{2}
\end{array}\right]\cdots\left[\begin{array}{c}
k_{p-1}\\
k_{p}
\end{array}\right].
\end{split}
\end{equation}
Similarly to the~trefoil case, for simplicity we omit the~prefactor $x^{p(\frac{\log a}{\hbar}-1)}$, which however has to be restored when using the~Weyl symmetry \eqref{Weyl_symmetry}. Summing up, the~closed form expression for the~$a$-deformed $F_K$ invariant for arbitrary $(2,2p+1)$ torus knot is given by
\begin{equation}\label{eq:F_K for torus knots}
\begin{split}F_{T^{(2,2p+1)}}(x,a,q)=\sum_{0\leq k_{p}\leq\ldots\leq k_{1}} & x^{2(k_{1}+\ldots+k_{p})-k_{1}}q^{(k_{1}+k_{2}+\ldots+k_{p})-\sum_{i=2}^{p}k_{i-1}k_{i}}\\
\times & \frac{(aq^{-1};q)_{k_{1}}(x;q^{-1})_{k_{1}}}{(q;q)_{k_{1}}}\left[\begin{array}{c}
k_{1}\\
k_{2}
\end{array}\right]\cdots\left[\begin{array}{c}
k_{p-1}\\
k_{p}
\end{array}\right].
\end{split}
\end{equation}

\subsection{Figure-eight knot}

Unfortunately, the~strategy devised above does not work for all knots. For example, the~(reduced) HOMFLY-PT polynomial for the~figure-eight knot is given in \cite{FGS} as
\begin{equation}
    P_r(4_1;a,q) = \sum_{k=0}^{\infty} (-1)^ka^{-k}q^{-k(k-3)/2}\frac{(aq^{-1};q)_k}{(q; q)_k}(q^{-r};q)_k(aq^{r};q)_k.
\end{equation}
Making the~substitution $x = q^r$, we get
\begin{equation}
    P_r(4_1;a,q) = \sum_{k=0}^{\infty} (-1)^ka^{-k}q^{-k(k-3)/2}\frac{(aq^{-1};q)_k}{(q; q)_k}(x^{-1};q)_k(ax;q)_k,
\end{equation}
but this expression does not give a~well-defined power series in $x$. 

Instead, we can recursively compute $F_{4_1}$ up to any desired order by noting that it should be annihilated by the~quantum $a$-deformed $A$-polynomial \cite{FGS}:
\begin{equation}
    \hat{A}_{4_1}(\hat{x},\hat{y},a,q) = a_0 + a_1 \hat{y} + a_2 \hat{y}^2 + a_3 \hat{y}^3
\end{equation}
where\footnote{Again we rescale $\hat{x}$ by $q$, $\hat{y}$ by $a/q$ and remove common factors of $a, q$. The~$A$-polynomial we use corresponds to the~reduced normalisation.}
\begin{align*}
    \begin{split}
        a_0 & = -\frac{(1-q\hat{x})(1-q^2\hat{x})(1-aq^4\hat{x}^2)(1-aq^5\hat{x}^2)}{q(1-aq\hat{x})(1-aq^2\hat{x})(1-aq\hat{x}^2)(1-aq^2\hat{x}^2)},
        \\ a_1 & = \frac{(1-q^2\hat{x})(1-aq^5\hat{x}^2)}{q^4\hat{x}^2(1-aq\hat{x})(1-aq^2\hat{x})(1-aq\hat{x}^2)}
        \\ & \quad \times \big(-1+2q^2\hat{x}+aq(1-q-q^2+q^3)\hat{x}^2+aq^3(-1+q+q^2-q^3)\hat{x}^3-2a^2q^5\hat{x}^4+a^2q^7\hat{x}^5\big),
        \\ a_2 & = \frac{(1-aq^4\hat{x}^2)}{q^4\hat{x}^2(1-aq^2\hat{x})(1-aq^2\hat{x}^2)} 
        \\ & \quad \times        
        \big(1-2aq\hat{x}-aq^2(1-q)^2(1+q)\hat{x}^2+a^2q^3(1-q-q^2+q^3)\hat{x}^3+2a^2q^7\hat{x}^4-a^3q^8\hat{x}^5\big),
        \\ a_3 & = \frac{a^2}{q}.
    \end{split}
\end{align*}
We take the~ansatz\footnote{The prefactor $x^{\frac{\log a}{\hbar}-1}$ is omitted.}
\[
    F_{4_1}(x, a, q) = \sum_{k = 0}^{\infty} f_k(a, q) x^k,
\]
and then use $\hat{A}_{4_1}$ to recursively solve for the~$f_k$. This leaves us with one free variable, $f_0$, which we set to be $1$ for now.\footnote{In general, we cannot simply set $f_0=1$; it should be determined by means other than recursion.} 
%in line with \eqref{eq:F_3_1} and \eqref{eq:F_K for torus knots}. 
The~first few terms are
\begin{align}
    \begin{split}
        f_0 & = 1,
        \\ f_1 & = -\frac{3(a-q)}{(1-q)},
        \\ f_2 & = -\frac{(a-q)(1-2a+6q-6aq+2q^2-aq^2)}{(1-q)(1-q^2)},
        \\ f_3 & = -\frac{(1-a)(a-q)(2+3q-5aq+11q^2-6aq^2+6q^3-11aq^3+5q^4-3aq^4-2aq^5}{(1-q)(1-q^2)(1-q^3)}.
    \end{split}
\end{align}
This looks relatively arcane but there is another form which makes things much clearer. Denoting
\[
    (a)^{(n)} = \prod_{i = 1}^n \frac{(a-q^i)}{(1-q^i)} = \frac{a^n(a^{-1}q;q)_n}{(q;q)_n}
\]
we find that we can write our functions as follows:
\begin{align}
    \begin{split}\label{eq:f_k for figure-eight}
        f_0 & = (a)^{(0)} = 1,
        \\ f_1 & = -3(a)^{(1)},
        \\ f_2 & =  - (1+6q+q^2)(a)^{(1)} + (2+6q+q^2)(a)^2,
        \\ f_3 & = - \Big(2+3q+11q^2+3q^3+2q^4\Big)(a)^{(1)} + \Big(2+8q + 17q^2 + 14q^3 + 5q^4 + 2q^5\Big)(a)^{(2)}
        \\ & \quad  -\Big(5q +6q^2 + 11q^3 + 3q^4 + 2q^5\Big)(a)^{(3)}.
    \end{split}
\end{align}

Whether we can solve the~recursion uniquely or not is determined by the~following proposition:
\begin{prop}
    Suppose that the~quantum $A$-polynomial is properly normalised so that we expect a~solution $f(x,a,q)$ to the~equation
    \begin{equation}\label{eq:qdiffeq}
        \hat{A}(\hat{x},\hat{y},a,q)f(x,a,q) = 0
    \end{equation}
    of the~form
    \begin{equation}\label{eq:fansatz}
        f(x,a,q) = c_0 + c_1(a,q) x + c_2(a,q) x^2 + \ldots
    \end{equation}
    with $c_0 = 1$. 
    Let us write the~quantum $A$-polynomial as follows: 
    \[
        \hat{A}(\hat{x},\hat{y},a,q) = \sum_{j=0}^{d}x^j b_j(\hat{y},a,q).
    \]
    Then the~equation \eqref{eq:qdiffeq} has a~unique solution of the~form \eqref{eq:fansatz} if and only if 
    \begin{equation}\label{eq:criterion}
        b_0(1,a,q) = 0
%    \end{equation}
\qquad    and \qquad 
%    \begin{equation}
        b_0(q^j,a,q) \neq 0
    \end{equation}
    for every $j \in \mathbb{Z}_+$. 
    If these conditions are satisfied, then the~unique solution is given recursively by
    \begin{equation}
        c_j = -\frac{1}{b_0(q^j,a,q)}\sum_{k=0}^{j-1}b_{j-k}(q^k,a,q)c_k
    \end{equation}
    for each $j \in \mathbb{Z}_+$.
\end{prop}
For the~figure-eight knot, $b_0 = y-1$, and this proves that there exists a~unique solution. This also shows why in this example $f_n$ multiplied by $(q;q)_n$ is a~polynomial.

\subsection{\texorpdfstring{$5_2$}{52} knot}\label{ssec:5_2}
Unfortunately, there are many knots for which \eqref{eq:criterion} is not satisfied (it seems that non-fiberedness is correlated with the~non-uniqueness). Twist knots $K_n$ with $|n|>1$ are good examples. For $K_n$, $b_0(y,a,q)$ has a~factor of $\prod_{j=0}^{|n|-1}(y-q^j)$, meaning that the~first $n$~coefficients, $f_0, f_1, \ldots, f_{n-1}$ are free parameters and cannot be determined by just solving the~recursion. Below we study the~example of $K_2 = 5_2$ in detail to illustrate this point. 

We find that, although the~first two coefficients $f_0, f_1$ seem to be free parameters, we can do better than that; in particular, by imposing non-singularity condition for $F_K(x,e^{N\hbar},e^{\hbar})$ in the~limit $\hbar \rightarrow 0$, the~term $f_1$ is determined by $f_0$. Schematically, 
\[(\hat{A}_K F_K = 0)\;+\;(\text{the non-singularity condition})\Rightarrow \text{unique solution, up to an~overall factor}.\]
The $5_2$ knot is just an~example, and we conjecture that this procedure works for every knot. In terms of the~curve of the~$A$-polynomial $\{A_K=0\}$, this means that we expect a~unique wave function once a~branch of the~curve near near $x=-\infty$ has been specified. We also note that the~appearance of many branches of the~curve of the~$A$-polynomial indicates that the~form $\log x\; d(\log y)$ is singular along the~curve.   

The (reduced) quantum $a$-deformed $A$-polynomial for the~$5_2$ knot can be found in \cite{NRZS, FGSS}. After aligning with the~conventions we are using, it is given by
\begin{equation}
\hat{A}_{5_2}(\hat{x},\hat{y},a,q) = a_0 +a_1\hat{y} +a_2\hat{y}^2 +a_3\hat{y}^3 +a_4\hat{y}^4,
\end{equation}
where
\begin{align*}
    a_0 &= -a q^{12} \hat{x}^7 (q \hat{x}-1) \left(q^2 \hat{x}-1\right) \left(q^3 \hat{x}-1\right) \left(a q^5 \hat{x}^2-1\right) \left(a q^6 \hat{x}^2-1\right) \left(a q^7 \hat{x}^2-1\right),\\
    a_1 &= q^6 \hat{x}^2 \left(q^2 \hat{x}-1\right) \left(q^3 \hat{x}-1\right) \left(a q^2 \hat{x}^2-1\right) \left(a q^6 \hat{x}^2-1\right) \left(a q^7 \hat{x}^2-1\right) \left(a^3 q^9 \hat{x}^6+a^3 q^8 \hat{x}^6\right.\\
    &\quad-3 a^2 q^8 \hat{x}^5-a^2 q^8 \hat{x}^4-a^2 q^7 \hat{x}^5-a^2 q^7 \hat{x}^4+a^2 q^6 \hat{x}^4-a^2 q^4 \hat{x}^4-a^2 q^3 \hat{x}^4+a q^8 \hat{x}^4+a q^7 \hat{x}^4\\
    &\quad+2 a~q^7 \hat{x}^3+a q^6 \hat{x}^4-a q^5 \hat{x}^3-a q^5 \hat{x}^2-a q^4 \hat{x}^3+2 a~q^3 \hat{x}^3+a q^3 \hat{x}^2+a q^2 \hat{x}^2-a q \hat{x}^2+q^4 \hat{x}^2\\
    &\quad\left.-2 q^2 \hat{x}+1\right),\\
%\end{align*}    
%\begin{align*}
    a_2 &= -q \left(q^3 \hat{x}-1\right) (a q \hat{x}-1) \left(a q \hat{x}^2-1\right) \left(a q^4 \hat{x}^2-1\right) \left(a q^7 \hat{x}^2-1\right) \left(a^4 q^{16} \hat{x}^8-2 a^3 q^{15} \hat{x}^7\right.\\
    &\quad-a^3 q^{14} \hat{x}^7-a^3 q^{14} \hat{x}^6-a^3 q^{13} \hat{x}^6-a^3 q^{11} \hat{x}^6-a^3 q^{10} \hat{x}^6+2 a^2 q^{14} \hat{x}^6+3 a^2 q^{13} \hat{x}^6+2 a^2 q^{13} \hat{x}^5\\
    &\quad+a^2 q^{12} \hat{x}^5-2 a^2 q^{11} \hat{x}^5+a^2 q^{11} \hat{x}^4-a^2 q^{10} \hat{x}^5+2 a^2 q^9 \hat{x}^5+a^2 q^9 \hat{x}^4+a^2 q^8 \hat{x}^5+2 a^2 q^8 \hat{x}^4\\
    &\quad+a^2 q^7 \hat{x}^4+a^2 q^5 \hat{x}^4-a q^{13} \hat{x}^5-a q^{12} \hat{x}^5-2 a~q^{12} \hat{x}^4-a q^{11} \hat{x}^5-2 a~q^{11} \hat{x}^4+a q^{10} \hat{x}^4+2 a~q^9 \hat{x}^4\\
    &\quad+2 a~q^9 \hat{x}^3-a q^8 \hat{x}^4+a q^8 \hat{x}^3-2 a~q^7 \hat{x}^4-2 a~q^7 \hat{x}^3-a q^6 \hat{x}^3-a q^6 \hat{x}^2+2 a~q^5 \hat{x}^3-a q^5 \hat{x}^2\\
    &\quad\left.+a q^4 \hat{x}^3-a q^3 \hat{x}^2-a q^2 \hat{x}^2-q^9 \hat{x}^3-q^8 \hat{x}^3-q^7 \hat{x}^3+2 q^6 \hat{x}^2+3 q^5 \hat{x}^2-2 q^3 \hat{x}-q^2 \hat{x}+1\right),\\
    a_3 &= (a q \hat{x}-1) \left(a q^2 \hat{x}-1\right) \left(a q \hat{x}^2-1\right) \left(a q^2 \hat{x}^2-1\right) \left(a q^6 \hat{x}^2-1\right) \left(a^3 q^{16} \hat{x}^6-2 a^2 q^{14} \hat{x}^5-a^2 q^{13} \hat{x}^4\right.\\
    &\quad+a^2 q^{11} \hat{x}^4+a^2 q^{10} \hat{x}^4-a^2 q^9 \hat{x}^4+a q^{12} \hat{x}^4+2 a~q^{11} \hat{x}^3-a q^9 \hat{x}^3-a q^8 \hat{x}^3-a q^8 \hat{x}^2+2 a~q^7 \hat{x}^3\\
    &\quad\left.-a q^7 \hat{x}^2+a q^6 \hat{x}^2-a q^4 \hat{x}^2-a q^3 \hat{x}^2+q^8 \hat{x}^2+q^7 \hat{x}^2+q^6 \hat{x}^2-3 q^4 \hat{x}-q^3 \hat{x}+q+1\right),\\
    a_4 &= (a q \hat{x}-1) \left(a q^2 \hat{x}-1\right) \left(a q^3 \hat{x}-1\right) \left(a q \hat{x}^2-1\right) \left(a q^2 \hat{x}^2-1\right) \left(a q^3 \hat{x}^2-1\right).
\end{align*}
Solving the~recursion, we find that the~first two coefficients, $f_0$ and $f_1$, determine all the~others. That is,\footnote{The prefactor $x^{\frac{\log a}{\hbar}-1}$ is omitted.}
\begin{equation}
    F_{5_2}(x,a,q) = \sum_{j\geq 0}f_j(a,q)x^j,
\end{equation}
where $f_j$ with $j\geq 2$ are $\mathbb{Q}(a,q)$-linear combinations of $f_0$ and $f_1$. For instance, 
\[
f_2 = -\frac{\left(a^2+a \left(q^2-4 q-2\right)+q \left(-q^2+3 q+2\right)\right)}{(q-1)^2 (q+1)}f_0 +\frac{(a q+a-3 q-1)}{q^2-1}f_1.
\]
Although at this point it may seem like $f_0$ and $f_1$ are free parameters, we have additional conditions to impose, namely that $F_K(x,q^N,q)$ is non-singular in the~semiclassical limit $\hbar \rightarrow 0$. Note that this non-singularity condition is weaker than imposing the~explicit limit~\eqref{eq:mixedlimit}, but as we will see, powers of Alexander polynomial automatically pop up just from this non-singularity condition. This non-singularity property imposes lots of conditions on the~perturbative coefficients of $f_0$ and $f_1$, and in particular it determines the~entire perturbative series of the~ratio $f_1/f_0$: 
\begin{align}
    \frac{f_1}{f_0}(a=e^{N\hbar},q=e^{\hbar}) &= \frac{3}{2}(N-1)\nonumber\\
    &\quad+\frac{5}{8}N(N-1)\hbar\nonumber\\
    &\quad+\frac{3}{16}N(N-1)(2N-1)\frac{\hbar^2}{2!}\\
    &\quad+\frac{1}{64}N(N-1)(17N^2-17N-3)\frac{\hbar^3}{3!}\nonumber\\
    &\quad+\frac{1}{320}N(N-1)(66N^3-99N^2+11N+41)\frac{\hbar^4}{4!}\nonumber\\
    &\quad \quad \vdots\nonumber
\end{align}
Plugging this back in and setting $\lim_{q\rightarrow 1}f_0(q^N,q) = 2^{1-N}$, we see that the~expected properties \eqref{eq:Alexanderlimit}-\eqref{eq:mixedlimit} in Conjecture \ref{$a$-deformed $F_K$} hold! That is, when $a=1$, we have $f_0(q^0,q)=2$, $f_1(q^0,q)=-3$, and 
\begin{equation}
    F_{5_2}(x,q^0,q) = 2-3x+2x^2,
\end{equation}
whereas for $a=q$ we get 
\begin{equation}
    F_{5_2}(x,q^1,q) = 1.
\end{equation}
Similarly, when $a=q^N$ ($N$ not necessarily an~integer), we have
\begin{equation}
    \lim_{q\rightarrow 1} F_{5_2}(x,q^N,q) = (2-3x+2x^2)^{1-N}.
\end{equation}

Expressing the~series in terms of $\hbar$ and $N\hbar$ instead, we get
\begin{align}
    (q-1)\frac{f_1}{f_0}(a,q)\bigg\vert_{a=e^{N\hbar},q=e^{\hbar}} &= \qty(3(N\hbar/2) +5\frac{(N\hbar/2)^2}{2!} +9\frac{(N\hbar/2)^3}{3!} +17\frac{(N\hbar/2)^4}{4!} +\ldots)\nonumber\\
    &\quad +\qty(-\frac{3}{2} +\frac{1}{4}(N\hbar/2) +\frac{1}{4}\frac{(N\hbar/2)^2}{2!} +\frac{1}{4}\frac{(N\hbar/2)^3}{3!} +\ldots)\hbar\nonumber\\
    &\quad+\qty(-\frac{3}{2}+\frac{1}{8}(N\hbar/2)+\frac{1}{8}\frac{(N\hbar/2)^3}{3!}+\frac{1}{8}\frac{(N\hbar/2)^5}{5!}+\ldots)\frac{\hbar^2}{2!}\\
    &\quad+\qty(-\frac{3}{2}+\frac{5}{32}(N\hbar/2)+\frac{1}{16}\frac{(N\hbar/2)^2}{2!} + \frac{13}{32}\frac{(N\hbar/2)^3}{3!}+\ldots)\frac{\hbar^3}{3!}\nonumber\\
    &\quad \quad \vdots\nonumber
\end{align}
Resumming this perturbative series into an~expression in $a$ and $q$, we see the~first few terms of $(q-1)f_1/f_0(a,q)$:
\begin{align}
    (q-1)\frac{f_1}{f_0}(a,q) &= a+a^{1/2}-2\nonumber\\
    &\quad+ \qty(\frac{a^{1/2}-7}{4})\hbar\nonumber\\
    &\quad+ \qty(\frac{a^{1/2}-24-a^{-1/2}}{16})\frac{\hbar^2}{2!}\nonumber\\
    &\quad+ \qty(\frac{a^{1/2}-82-21a^{-1/2}+6a^{-1}}{64})\frac{\hbar^3}{3!}+ \ldots\nonumber\\
    & = a~+ \left(1 + (\hbar/4) + \frac{(\hbar/4)^2}{2!} + \frac{(\hbar/4)^3}{3!} + \ldots \right)a^{\frac{1}{2}} \\ 
    & \quad + \left(-2-7(\hbar/4)-24\frac{(\hbar/4)^2}{2!}-82\frac{(\hbar/4)^3}{3!} - \ldots\right)\nonumber \\
    & \quad + \left(-1\frac{(\hbar/4)^2}{2!} - 21\frac{(\hbar/4)^3}{3!} -  262\frac{(\hbar/4)^4}{4!} - \ldots\right)a^{-\frac{1}{2}}+\ldots\nonumber \\
    & = a~+ a^{\frac{1}{2}}q^{\frac{1}{4}} + \ldots.\nonumber
\end{align}
It is an~interesting problem to resum further terms into a~series in $q$ and $a$. 
%{\shp{Can we explain why non-integer powers appear here? Is it somehow ``equivalent'' to a~series only with integer powers?}}
%Another subtle point is that in general we cannot simply set $f_0=1$ as this would violate \eqref{eq:mixedlimit} in Conjecture \ref{$a$-deformed $F_K$}. 

\section{A \texorpdfstring{$t$}{t}-deformation of \texorpdfstring{$F_K$}{FK}}\label{sec:t-deformation}
In the~previous section we found that the~$a$-deformed $F_{K}$ invariants for $(2,2p+1)$ torus knots can be derived from the~HOMFLY-PT polynomials. Since the~latter admit a~categorification \cite{Kho}, which was quite unexpected in the~mathematical literature and emerged from physics \cite{GSV,DGR}, we can follow this path and propose ($a,t$)-deformed $F_{K}$ invariants based on the~superpolynomials. These ($a, t$)-deformed invariants can also be computed term by term using the~super-$A$-polynomial introduced in~\cite{FGSA,FGS}.

\subsection{\texorpdfstring{$(2,2p+1)$}{(2,2p+1)} torus knots}

We start from the~unknot corresponding to $p=0$. The~formula for the~fully unreduced superpolynomial is given by \cite{FGS}
\begin{equation}
\bar{\mathcal{P}}_{r}(0_{1};a,q,t)=a^{-\frac{r}{2}}q^{\frac{r}{2}}(-t)^{-\frac{3r}{2}}\frac{(-at^{3};q)_{r}}{(q;q)_{r}}=\frac{(-at^{3};q)_{\infty}(q^{r+1};q)_{\infty}}{(-aq^{r}t^{3};q)_{\infty}(q;q)_{\infty}}.
\end{equation}
After substituting $q^{r}=x$ and dropping the~prefactors, we obtain the~$(a,t)$-deformed $F_K$ invariant in the~unreduced normalisation:
\begin{equation}
F^{\textrm{unred.}}_{0_{1}}(x,a,q,t)=\frac{(xq;q)_{\infty}}{(-xat^{3};q)_{\infty}}.\label{eq:t-deformed F_0_1}
\end{equation}
It is annihilated by the~quantum super-$A$-polynomial
\begin{equation}
\hat{A}_{0_{1}}(\hat{x},\hat{y},a,q,t)=(1+at^{3}\hat{x})-(1-q\hat{x})\hat{y},
\end{equation}
which, just as in the~previous section, agrees with \cite{FGS} after taking into account the~changes of conventions. Moreover, we can see that (\ref{eq:t-deformed F_0_1}) matches (\ref{F_0_1}) for $t=-1$.

For $(2,2p+1)$ torus knots with $p\geq1$, we use the~formula for reduced superpolynomials from \cite{FGSS}:
\begin{equation}
\begin{split}
\mathcal{P}_{r}(T^{(2,2p+1)};a,q,t)=\sum_{0\leq k_{p}\leq\ldots\leq k_{1}\leq k_{0}=r} & q^{(2r+1)(k_{1}+k_{2}+\ldots+k_{p})-\sum_{i=1}^{p}k_{i-1}k_{i}}t^{2(k_{1}+\ldots+k_{p})} \\
\times a^{pr}q^{-pr}& \frac{(q^{r};q^{-1})_{k_{1}}(-aq^{-1}t;q)_{k_{1}}}{(q;q)_{k_{1}}}\left[\begin{array}{c}
k_{1}\\
k_{2}
\end{array}\right]\cdots\left[\begin{array}{c}
k_{p-1}\\
k_{p}
\end{array}\right].
\end{split}
\end{equation}
Substituting $q^{r}=x$ and omitting the~prefactor $x^{p(\frac{\log a}{\hbar}-1)}$ (in analogy to Section \ref{sec:Torus knots}), we obtain the~closed form expression for the~$(a,t)$-deformed $F_{T^{(2,2p+1)}}$ in the~reduced normalisation:
\begin{equation}\label{eq:t-deformed F_K for torus knots}
\begin{split}
F_{T^{(2,2p+1)}}(x,a,q,t)=\sum_{0\leq k_{p}\leq\ldots\leq k_{1}} & x^{2(k_{1}+\ldots+k_{p})-k_{1}}q^{(k_{1}+k_{2}+\ldots+k_{p})-\sum_{i=2}^{p}k_{i-1}k_{i}}t^{2(k_{1}+\ldots+k_{p})}\\
\times & \frac{(-aq^{-1}t;q)_{k_{1}}(x;q^{-1})_{k_{1}}}{(q;q)_{k_{1}}}\left[\begin{array}{c}
k_{1}\\
k_{2}
\end{array}\right]\cdots\left[\begin{array}{c}
k_{p-1}\\
k_{p}
\end{array}\right].
\end{split}
\end{equation}
One can easily check that this formula agrees with (\ref{eq:F_K for torus knots}) for $t=-1$.

\subsection{Figure-eight knot}

Similarly, we can recursively solve for the~$t$-deformation of $F_{4_1}$ using the~super-$A$-polynomial. From \cite{FGS}, we find the~coefficients of the~super-$A$-polynomial $\hat{A}_{4_1}(\hat{x},\hat{y},a,q,t) = a_0 + a_1 \hat{y} + a_2 \hat{y}^2 + a_3 \hat{y}$ to be:\footnote{In this refined case we rescale $\hat{y}$ by $-at/q$, $\hat{x}$ by $q$ and then we remove common factors of $a, q, t$}
\begin{align}
    \begin{split}
        a_0 & = \frac{t^3(1-q\hat{x})(1-q^2\hat{x})(1+a q^4 t^3 \hat{x}^2)(1+aq^5t^3\hat{x}^2)}{q(1+aqt^3\hat{x})(1+aq^2t^3\hat{x})(1+aq^2t^3\hat{x}^2)(1+aqt^3\hat{x}^2)},
        \\ a_1 & = \frac{(1-q^2\hat{x})(1+aq^5t^3\hat{x}^2)}{q^4\hat{x}^2(1+aqt^3\hat{x})(1+aq^2t^3\hat{x})(1+aqt^3\hat{x}^2)},
        \\ & \quad \quad  \times\Big(1+q^2t(1-t)\hat{x}+aqt^3(1+q^3+qt+q^2t)\hat{x}^2-aq^3t^4(q+q^2+t+q^3t)\hat{x}^3
        \\ & \quad \quad \quad \quad +a^2q^5t^6(1-t)\hat{x}^4-a^2q^7t^8\hat{x}^5\Big),
        \\ a_2 & = -\frac{(1+aq^4t^3\hat{x}^2)}{q^4\hat{x}^2(1+aq^2t^3\hat{x})(1+aq^2t^3\hat{x}^2)}
        \\ & \quad \quad \times\Big(1+aqt(1-t)\hat{x} +aq^2t^2(q+q^2+t+q^3t)\hat{x}^2 + a^2q^3t^4(1+q^3+qt+q^2t)\hat{x}^3
        \\ & \quad \quad \quad \quad - a^2q^7t^5(1-t)\hat{x}^4 + a^3q^8t^7\hat{x}^5\Big),
        \\ a_3 & = -\frac{a^2t^3}{q}.
% Old version with x->qx applied, before cleaning
%        a_0 & = \frac{t^3(1-xq)(1-qxq)(1+a q^2 t^3 x^2q^2)(1+aq^3t^3x^2q^2)}{(1+at^3xq)(1+aqt^3xq)(1+at^3x^2q^2)(q+at^3x^2q^2)}
%        \\ a_1 & = \frac{(1-qxq)(1+aq^3t^3x^2q^2)}{qx^2(1+at^3xq)(1+aqt^3xq)(q+at^3x^2q^2)}
%        \\ & \quad \quad \quad \Big(1+qt(1-t)xq+at^3(q^{-1}+q^2+t+qt)x^2q^2-at^4(q+q^2+t+q^3t)x^3q^3
%        \\ & \quad \quad \quad \quad +a^2qt^6(1-t)x^4q^4-a^2q^2t^8x^5q^5\Big)
%        \\ a_2 & = -\frac{(1+aq^2t^3x^2q^2)}{q^2x^2q^2(1+aqt^3xq)(1+at^3x^2q^2)}
%        \\ & \quad \quad \quad \Big(1+at(1-t)xq +at^2(q+q^2+t+q^3t)x^2q^2 + a^2t^4(1+q^3+qt+q^2t)x^3q^3
%        \\ & \quad \quad \quad \quad - a^2q^3t^5(1-t)x^4q^4 + a^3q^3t^7x^5q^5\Big)
%        \\ a_3 & = -\frac{a^2t^3}{q}        
    \end{split}
\end{align}
Solving this recurrence relation as before, with the ansatz
\[
    F_{4_1}(x, a, q, t) = \sum_{k=0}^{\infty} f_k(a, q, t)x^k,
\]
we find that the~first few $f_k$ are given by 
\begin{align}
    \begin{split}
        f_0 & = (-at)^{(0)} = 1,
        \\ f_1 & = -(1 - t + t^2)(-at)^{(1)},
        \\ f_2 & = -\left(-q+qt+q^2t-t^2-2qt^2+t^3+qt^3-qt^4\right)(-at)^{(2)}
        \\ & \quad \quad \quad + \left(-q+qt+q^2t-2qt^2+t^3+qt^3-qt^4\right)(-at)^{(1)},
        \\ f_3 & = -q\Big(q^2-q^2 t-q^3 t-q^4 t+t^2+q t^2+2 q^2 t^2+q^3 t^2+q^4 t^2-t^3
        \\ & \quad \quad \quad \quad - 2 q t^3 - 3 q^2 t^3-q^3 t^3 2 t^4+2 q t^4+2 q^2 t^4-t^5-q t^5-q^2 t^5+q^2 t^6\Big)(-at)^{(3)}
        \\ & \quad \quad +\Big(q^2 + q^3 - q^2 t - 2 q^3 t - 2 q^4 t - q^5 t + q t^2 + 3 q^2 t^2 + 3 q^3 t^2
        \\ & \quad \quad \quad \quad  + 2 q^4 t^2 + q^5 t^2 - 2 q t^3 - 5 q^2 t^3 - 4 q^3 t^3 - q^4 t^3 + t^4 + 3 q t^4
        \\ & \quad \quad \quad \quad + 4 q^2 t^4 + 2 q^3 t^4 - t^5 - 2 q t^5 - 2 q^2 t^5 - q^3 t^5 + q^2 t^6 + q^3 t^6\Big)(-at)^{(2)}
        \\ & \quad \quad -\Big(q^2-q^2 t-q^3 t-q^4 t+2 q^2 t^2+q^3 t^2+q^4 t^2-q t^3
        \\ & \quad \quad \quad \quad -3 q^2 t^3-q^3 t^3+t^4+q t^4+2 q^2 t^4-t^5-q t^5-q^2 t^5+q^2 t^6\Big)(-at)^{(1)},
    \end{split}
\end{align}
where $(-at)^{(n)} = \prod_{i = 1}^n \frac{(-at-q^i)}{(1-q^i)}$. It is easy to see that for $t = -1$ we recover our previous result \eqref{eq:f_k for figure-eight}.

\subsection{\texorpdfstring{$t$}{t}-deformed Alexander and ADO polynomials}
Curiously, upon specialisation $a = -t^{-1}$, the~reduced $(a,t)$-deformed $F_K$ invariant becomes a~version of $t$-deformed Alexander polynomial, which we will call $\Delta_K(x,t)$. For instance, we have
\begin{equation} \label{Alexander Polynomials}
\begin{split}
    \Delta_{3^r_1}(x,t)&=-t^{-1}x^{-1}+t^{-1}-tx,
    \\ \Delta_{4_1}(x,t)&=t^{-1}x^{-1}+(-t^{-1}+1-t)+tx,
\end{split}
\end{equation}
for the~right-handed trefoil and figure-eight knot. Note that the~Weyl symmetry is $t$-deformed as well:
\begin{equation}
    \Delta_K(x^{-1},t) = \Delta_K(t^{-2}x,t).
\end{equation}
Naturally, these $t$-deformed Alexander polynomials can be obtained from the~usual superpolynomial $\mathcal{P}_K(a,q,t)$: 
\begin{equation}
    \Delta_K(x,t) = \mathcal{P}_K(-t^{-1},x,t).
\end{equation}
We experimentally found the~$t$-deformed version of the~conjectural equations \eqref{eq:Alexanderlimit}-\eqref{eq:mixedlimit}: 
\begin{align}
    F_K(x,-t^{-1},q,t) & = \Delta_K(x,t), \\
    F_K(x,-t^{-1}q,q,t) & = 1, \\
    \label{q to 1 limit}
    \lim_{q\rightarrow 1}F_K(x,-t^{-1}q^{N},q,t) & = \frac{1}{\Delta_K(x,t)^{N-1}}.
\end{align}

In a~recent work \cite{GHNPPS}, certain connections between $\widehat{Z}$ and non-semisimple modular tensor categories were observed. In particular, in Conjecture 3 of that paper, it was conjectured that ADO polynomials can be obtained as limits of $F_K$ as $q$ approaches roots of unity $\zeta_p = e^{2\pi i/p}$, up to a~factor determined by the~Alexander polynomial $\Delta_K(x)$. 
Since certain specialisations of the~$(a,t)$-deformed $F_K$ lead to $t$-deformed Alexander polynomials $\Delta_K(x,t)$, it is tempting to use them to study certain limits of $(a,t)$-deformed $F_K$ as $q$ approaches roots of unity. Below we give a~$t$-deformed version of Conjecture 3 of \cite{GHNPPS}, where the~role of Alexander polynomial is replaced by that of the~$t$-deformed one. 
\begin{conj}
    The~limit
    \begin{equation}
    \mathrm{ADO}_K(p;x,t):=\Delta_K(x^p,-(-t)^p)\lim_{q\rightarrow e^{2\pi i/p}}F_K(x,-t^{-1}q^2,q,t)
    \end{equation}
    is a~polynomial, and when $t=-1$, it is the~usual $p$-th ADO polynomial of $K$ for $SU(2)$. 

    More generally, the~limit 
    \begin{equation} \label{eq:tdefSUNADO}
    \mathrm{ADO}_K^{SU(N)}(p;x,t):=\Delta_K(x^p,-(-t)^p)^{N-1}\lim_{q\rightarrow e^{2\pi i/p}}F_K(x,-t^{-1}q^N,q,t)
    \end{equation}
    is a~polynomial and is a~$t$-deformation of the~symmetric version of the~$p$-th ADO polynomial of $K$ for $SU(N)$ (i.e.\ $P_K^{SU(N)}$ introduced in \cite{GHNPPS}, specialised to $x_1=qx$, $x_2=\cdots=x_{N-1} =q$). 
\end{conj}
Using the~data we provided in this section, we can explicitly compute the~$t$-deformed ADO polynomials for many knots. Below, in Tables \ref{tab:tdefADOtrefoil} and \ref{tab:tdefADOfigure8}, we summarise the~computation for the~trefoil and the~figure-eight knot. We have written in a~way that the~Weyl symmetry is manifest. Note that $\mathrm{ADO}_K(2;x,t)$ is always $\Delta_K(x,t)$. 
\begin{table}[ht]
    \centering
    \begin{tabular}{ccc}
         $p$ & & $\mathrm{ADO}_{3_1}(p;x,t)$ \\
         \hline\hline
         $1$ & & $1$\\
         $2$ & & $(-tx)+t^{-1}+(-tx)^{-1}$\\
         $3$ & & $(t\zeta_3x)^2+t^{-1}(t\zeta_3x)+(t^{-2}-\zeta_3)+t^{-1}(t\zeta_3x)^{-1}+(t\zeta_3x)^{-2}$\\
         %$4$ & & $(t\zeta_4x)^3+t^{-1}(t\zeta_4x)^2+(t^{-2}-\zeta_4)(t\zeta_4x)+((1-\zeta_4)t^{-1}+t^{-3})+(t^{-2}-\zeta_4)(t\zeta_4x)^{-1}+t^{-1}(t\zeta_4x)^{-2}+(t\zeta_4x)^{-3}$\\
         \hline\hline
    \end{tabular}
    \caption{$t$-deformed ADO polynomials for the~right-handed trefoil knot}
    \label{tab:tdefADOtrefoil}
\end{table}
\begin{table}[ht]
    \centering
    \begin{tabular}{ccc}
         $p$ & & $\mathrm{ADO}_{4_1}(p;x,t)$ \\
         \hline\hline
         $1$ & & $1$\\
         $2$ & & $-(-tx)+(-t+1-t^{-1})-(-tx)^{-1}$\\
         $3$ & & $(t\zeta_3 x)^2+(t-1+t^{-1})(t\zeta_3 x)+(t^2+\zeta_3 t+2+\zeta_3^{-1}t^{-1}+t^{-2})$\\%+(t-1+t^{-1})(t\zeta_3 x)^{-1}+(t\zeta_3 x)^{-2}$\\
         & &  $+(t-1+t^{-1})(t\zeta_3 x)^{-1}+(t\zeta_3 x)^{-2} $\\
         %$4$ & & $(t\zeta_4x)^3+(t-1+t^{-1})(t\zeta_4x)^2+(t^2+(\zeta_4-1)t+2-(\zeta_4+1)t^{-1}+t^{-2})(t\zeta_4x)+(-\zeta_4t^3+t^2-(\zeta_4+1)t+3\zeta_4+(1-\zeta_4)t^{-1}-t^{-2}-\zeta_4t^{-3})+\cdots$\\
         \hline\hline
    \end{tabular}
    \caption{$t$-deformed ADO polynomials for the~figure-eight knot}
    \label{tab:tdefADOfigure8}
\end{table}
\iffalse
\begin{table}[ht]
    \centering
    \begin{tabular}{ccc}
         $p$ & & $\mathrm{ADO}_{3_1}^{SU(3)}(p;x,t)$ \\
         \hline\hline
         $1$ & & $1$\\
         $2$ & & $t^2x^2-t^{-2}+t^{-2}x^{-2}$\\%may be off by a~sign
         $3$ & & $...$\\
         \hline\hline
    \end{tabular}
    \caption{$t$-deformed symmetric $SU(3)$-ADO polynomials for the~right-handed trefoil knot}
    \label{tab:tdefSU3ADOtrefoil}
\end{table}
\fi

Computations for the~first few values of $N$ suggest that the~$t$-deformed $SU(N)$-ADO polynomial has the~Weyl symmetry
\begin{equation}
    \mathrm{ADO}_K^{SU(N)}(p;x^{-1},t) = \mathrm{ADO}_K^{SU(N)}(p;\zeta_p^{-N}t^{-2}x,t).
\end{equation}
This in turn suggests the~Weyl symmetry of the~$(a,t)$-deformed $F_K$:
\begin{equation}
    F_K(x^{-1},-t^{-1}a,q,t) = F_K(t^{-2}a^{-1}x,-t^{-1}a,q,t),
\end{equation}
or simply
\begin{equation}
    F_K(x^{-1},a,q,t) = F_K(-t^{-3}a^{-1}x,a,q,t).
\end{equation}

Testing out different values of $p$ and $N$ on small knots such as $3_1, 4_1$, and $5_1$, it appears that the $SU(N)$-ADO polynomials fall into patterns depending on $N \mod p$:
\begin{equation}
    \mathrm{ADO}^{SU(N + p)}_{K}(p; x, t) = \Delta_{K}(x^p, -t^p)^{p-1}\mathrm{ADO}^{SU(N)}_{K}(p; x, t),
\end{equation}
This, in particular, implies 
\iffalse
\begin{align}
    \begin{split}
        \mathrm{ADO}^{SU(2N)}_{K}(2; x, t) & = \Delta_{K}(x, t)\Delta_{K}(x^2, -t^2)^{N-1},
        \\ \mathrm{ADO}^{SU(2N + 1)}_{K}(2; x, t)  & = \Delta_{K}(x^2, -t^2)^N,
    \end{split}
\end{align}
as well as 
\fi
\begin{align}
    \begin{split}
        \mathrm{ADO}^{SU(pN)}_{K}(p; x, t) &= \Delta_{K}(x, t)\Delta_{K}(x^p, -(-t)^p)^{(p-1)N-1},
        \\ \mathrm{ADO}^{SU(pN+1)}_{K}(p; x, t) &= \Delta_{K}(x^p, -(-t)^p)^{(p-1)N}.
    \end{split}
\end{align}

Finally, we note that it is also possible to make sense of $SU(N)$-ADO invariant for generic~$N$. This gives rise to a~two variable series in $x,t$ with coefficients being functions of~$N$. 
\iffalse
In the~case of the~right-handed trefoil, it follows from equations \eqref{eq:t-deformed F_K for torus knots, Alexander Polynomials}\footnote{Note that equation \eqref{eq:t-deformed F_K for torus knots} is given for the~left handed trefoil, but it is easy to convert it to a~similar expression for the~right handed one.} that we will get a structure of the form 
\begin{align}
    \begin{split}
        \mathrm{ADO}_{3_1^r}(p; N, x, t) & = \Delta_{3_1^r}(x^p,-(-t)^p)^{N-1}\lim_{q\rightarrow e^{2\pi i/p}}F_{3_1^r}(x,-t^{-1}q^N,q,t)
        \\ & = \big(tx\big)^{(p - 1)(1 - N)} \sum_{i, j \in \mathbb{N}} f^p_{i,j}(N) \ x^it^j.
    \end{split}
\end{align}
When $p = 1$, it follows from equation \eqref{q to 1 limit} that this series is simply $1$, so we focus our attention on $p > 1$. 
\fi
The~first couple of terms in this series for the~right-handed trefoil are shown in Table \ref{tab:NtdefADOtrefoil}. 
\begin{table}[ht]
    \centering
    \begin{tabular}{ccc}
         $p$ & & $\big(tx\big)^{(p - 1)(N - 1)}\mathrm{ADO}_{3_1^r}^{SU(N)}(p;x,t)$ \\
         \hline\hline
         $2$ & & $\left(1 - \frac{1}{2}(1 + (-1)^{N})x + \frac{1}{4}(3 + (-1)^N - 2N)x^2 + O(x^3)\right)$\\
             & & $\quad + \left(\frac{1 + (-1)^N}{2}x^2 - \frac{1 + (-1)^N}{4}(-2 + N)x^4 + O(x^6)\right)t^2$ \\
             & & $\quad + \left(-\frac{3 + (-1)^N + 2N}{4}x^4 - \frac{1 + (-1)^N}{4}(-2 + N)x^5 + O(x^6)\right)t^4 + O(t^6)$ \\
         $3$ & & $\left(1 + \frac{1}{3}\left((\zeta_3 - 1) + ((\zeta^2_3 - 1))\zeta_3^{2N} \right)x + O(x^2)\right)$ \\
             & & $\quad + \left(\frac{\zeta_3^N}{3}\Big(1 - \zeta_3 + (1 - \zeta_3^2)\zeta_3^{N}\Big)x^2 + O(x^3)\right)t^2 + O(t^4)$ \\
         $4$ & & $\left(1 + \frac{1}{2}(i-1)(1+i^{N+1})x+\frac{1}{4}(i-1)(1-i^N)(i-i^N)x^2 + O(x^3)\right)$\\
         & & $\quad + \left(\frac{(-1)^N}{2}(1-i)(i+i^{3N})x^2 + O(x^3)\right)t^2 + O(t^4)$ \\
         \hline\hline
    \end{tabular}
    \caption{$N, t$-deformed ADO series for the~right-handed trefoil knot}
    \label{tab:NtdefADOtrefoil}
\end{table}

\subsection{Physical and geometric meaning of \texorpdfstring{$t$}{t}}

Physically, the~$q$-series invariants $\widehat{Z}$ and their variants $F_K$ for knot complements are generating functions of integer BPS invariants, cf. the~lower-right corner of Table~\ref{tab:enumerative}.
In particular, they are defined as graded traces over the~spaces of BPS states, $\mathcal{H}^{\text{BPS}}_{i,j,\beta}$, so that the~latter provide categorification of $\widehat{Z}$ and $F_K$.

In the~same way, the~$a$-dependent invariants studied in this paper encode the~graded dimensions of the~spaces of BPS states on the~{\it resolved} side of the~geometric transition \eqref{Mresolved}:
\begin{equation}
F_K (x,a,q) \; := \; \sum_{i,j,\beta} (-1)^i q^j a^{\beta} \, \text{rank} \mathcal{H}^{\text{BPS}}_{i,j,\beta} (X,L_K)
\end{equation}
Introducing a~new variable $t'$ and replacing $(-1)^i$ on the~right-hand side by $(t')^i$, we obtain the~Poincar\'e polynomial of $\mathcal{H}^{\text{BPS}} (X,L_K)$. In the~case of knots in $S^3$, this gives the~Poincar\'e polynomial of the~coloured HOMFLY-PT homology (a.k.a.\ the~coloured superpolynomial) such that, possibly up to a~simple change of variables, $t'=t$.

One motivation for the~present work is to gain access to $\mathcal{H}^{\text{BPS}}$ in the~case of 3-manifolds. Note that in this case the~Calabi-Yau geometry $X$ on the~resolved side \eqref{Mresolved} depends on the~choice of 3-manifold $Y$. Of course, for $Y = S^3$ with no knots in it, $X$ is just the~resolved conifold. Even in this case, the~space of BPS states has a~very rich structure \cite{GPV}, in particular it has the~right structure to produce spaces $\mathcal{H}^{\text{BPS}}_{SU(N)}$ on the~deformed side via spectral sequences with differentials $d_N$. These are the~same type of differentials that relate  e.g. Khovanov homology and its Lee deformation.
For other 3-manifolds the~spaces of BPS states on the~resolved side are not known explicitly. However, if one can identify $t'=t$ as in the~case of knot invariants, then the~computation of $t$-dependent $F_K (x,a,q,t)$ in this paper provides the~desired graded Poincar\'e polynomial of $\mathcal{H}^{\text{BPS}} (M_K)$ on the~resolved side. Whether this is true can be checked in a~number of ways.

For example, one can ask whether the~$t$-deformation $F_K (x,a,q,t)$ computed here has the~right structure to reproduce the~finite-rank $SU(N)$ version by taking cohomology with respect to the~differentials $d_N$, as in the~case of knots and as in the~case of $Y = S^3$.
In other words, the~differentials $d_N$ relate the~spaces of BPS states on the~resolved (HOMFLY-PT) side and on the~deformed $SU(N)$ side. Moreover, if we know $t$-deformed $\widehat{Z}$-invariants on both sides, we can simply check whether the~difference is of the~form $(1 + t^i q^j a^{\beta}) (\ldots)$ with particular $(i,j,\beta) = \deg (d_N)$ as in the~case of knots and $Y = S^3$.

From the~point of view of the~geometric interpretation of quiver nodes as basic holomorphic disks \cite{EKL1,EKL2}, %and the~form of the~refined quiver partition function 
it is natural to conjecture that the~counterpart of refined Chern-Simons theory has to do with distinguishing the~self-linking of the~boundary of a~basic disk from the~4-chain intersections in its interior. For bare curves (i.e.\ perturbed curves of positive symplectic area that are embedded) %, as explained in Ekholm-Shende, 
self-linking of the~boundary can be traded for 4-chain intersections \cite{ES}. For basic disks the~situation is different, as they should be considered with all their multiple covers. %this should not be the~case. 
Note here that when counting generalised holomorphic curves induced by degree $d$ covers of the~basic disk, the~boundary self-linking $\ell$ contributes quadratically~($q^{\ell d^{2}}$), whereas the~4-chain intersection $c$ contributes linearly~($q^{c d}$).
From the~point of view of knot conormals, one would guess that refined invariants are defined for (possibly singular) special Lagrangains associated to a~knot. Candidates for such Lagrangians are covers of the~unknot conormal branched along a~link.

Here we provide further evidence for how $4$-chain intersections contribute to refined curve counting. Consider a~basic holomorphic $\C {\bf P}^{1}$ in a~Calabi-Yau 3-fold and a~Lagrangnian brane~$L$ that moves so as to intersect the~basic sphere. Then the~basic sphere passes through the~Lagrangian, but a~new moduli space consisting of a~basic disk appears. The~difference between the~sphere before and after the~crossing with $L$ is that one $4$-chain intersection changes its sign, see Figure \ref{fig:refinedsphere}. 
\begin{figure}[htp]
	\centering
	\begin{tikzpicture}[scale=0.75]
    	\draw (0, 0) arc(0:360:2 and 0.75);
    	\draw[rotate=-10] (-2, 0.4) arc(90:-90:0.25 and 0.75);
    	\draw[rotate=-10,dashed] (-2, 0.4) arc(90:270:0.25 and 0.75);
    
    	\draw[gray] (0, -1.5) -- (1, -2.5) -- (1, 1) -- (0, 2);
    
    	\node at (-6, 3) {Before crossing:};
    	\node[gray] at (0.5, -2.5) {$L$};
    	\node[blue] at (0, 0.75) {$+$};
    	\draw[blue, fill] (-0.25, 0) circle[radius=1pt];
    	\node at (-2, -1.5) {$\mathbb{C}P^1_{+}$};
    
    	\node at (-6, -3) {After crossing:};
    
    	\draw (-4 + 0, 0 - 5.75) arc(30:330:2 and 0.75);
    	\draw[densely dashed] (-4 + 0, 0 - 5.75) arc(90:450:0.2 and 0.375);
    
    	\draw[gray] (-4.5 + 0, -1.5 - 6) -- (-4.5 + 1, -2.5 - 6) -- (-4.5 + 1, 1 - 6) -- (-4.5 + 0, 2 - 6);
    	\node[gray] at (-4, -2.5  - 6) {$L$};
    
    	\node at (-6, -1.5 - 6) {$\mathbb{D}$};
    
    	\draw (4 + 0, 0 - 6) arc(0:360:2 and 0.75);
    	\draw[rotate=-10] (5 + -2, 0.4 - 5.205) arc(90:-90:0.25 and 0.75);
    	\draw[rotate=-10,dashed] (5 + -2, 0.4 - 5.205) arc(90:270:0.25 and 0.75);
    
    	\draw[draw=white,double=gray,double distance=\pgflinewidth,ultra thick] (2.5 + 0, -1.5 - 6) -- (2.5 + 1, -2.5 - 6) -- (2.5 + 1, 1 - 6) -- (2.5 + 0, 2 - 6);
    
    	\node[gray] at (3, -2.5  - 6) {$L$};
    
    	\node[blue] at (0 + 4, 0.75 - 6) {$-$};
    	\draw[blue, fill] (-1 + 4.75, 0.25 - 6.25) circle[radius=1pt];
    	\node at (-2 + 3, -1.5 - 6) {$\mathbb{C}P^1_{-}$};
    \end{tikzpicture}
	\caption{Basic holomorphic spheres and disks for refined curve counting.}
	\label{fig:refinedsphere}
\end{figure}
Assume now that the~basic disk that appears has no self-linking, so that its partition function is
\[ 
\exp\left(-\sum_{d=1}^{\infty}\frac{a^{d}}{d(q^{d/2}-q^{-d/2})}\right).
\]         
According to \cite{AS}, the~refined partition function of $\C {\bf P}^{1}$ is
\[ 
\exp\left(\sum_{d=1}^{\infty}\frac{a^{d}}{d(q^{d/2}-q^{-d/2})(t^{d/2}-t^{-d/2})}\right).
\] 
This then indicates that we should count the~contribution to the~refined open string by 4-chain intersections, which would then give an~invariant refined partition function as follows. Write $\C {\bf P}^{1}_{-}$ and $\C {\bf P}^{1}_{+}$ for the~sphere before and after the~$L$ intersection, respectively, and $\mathbb{D}$~for the~disk. Then 
\begin{align}
    \Psi_{\C {\bf P}^{1}_+}\Psi_{\C {\bf P}^{1}_{-}}^{-1}&=\exp\left(-\sum_{d=1}^{\infty}\frac{t^{d/2}a^{d}}{d(q^{d/2}-q^{-d/2})(t^{d/2}-t^{-d/2})}+\sum_{d=1}^{\infty}\frac{t^{-d/2}a^{d}}{d(q^{d/2}-q^{-d/2})(t^{d/2}-t^{-d/2})}\right)\nonumber\\
&=\exp\left(-\sum_{d=1}^{\infty}\frac{a^{d}}{d(q^{d/2}-q^{-d/2})}\right) = \Psi_{\mathbb{D}}.
\end{align}

\section{Mirror knots and relations with quantum modularity}\label{sec:Quantum modularity}

One of the~main results of this work is that the~geometric realisation of $F_K$ in terms of curve counting as well as the~explicit examples we studied all suggest the~existence of an~$a$-deformation of two-variable series $F_K (x,q)$ for knot complements.

A simple but interesting corollary of this is that for every knot $K$ the~$a$-deformed functions $F_K (x,a,q)$ come in pairs,
\begin{equation}
F_K (x,a,q) 
\quad \xleftrightarrow[~]{\text{~~~mirror~~~}} \quad
F_{m(K)} (x,a,q),
\label{mirrorFF}
\end{equation}
where $m(K)$ denotes the~mirror knot. If the~knot $K$ is amphichiral, i.e.\ $K \simeq m(K)$, then the~above relation is simply an~equality. It becomes very interesting and highly nontrivial, though, when $K$ is {\it not} amphichiral.\footnote{Actually it is interesting for non-fibered amphichiral knots too, as it will give us a~$q$-series which is invariant under $q\leftrightarrow q^{-1}$.}

In order to explain this simple but important point, let us recall that under $K \mapsto m(K)$ the~coloured HOMFLY-PT polynomials behave as
\begin{equation}
P_{R} (m(K);a,q) \; = \; P_{R} (K;a^{-1},q^{-1})
\label{mirrorHOMFLY}
\end{equation}
for any colour $R$ (including the~symmetric ones, most relevant to us here). The~behaviour is consistent with (and can be derived from) a~similar behaviour in Chern-Simons theory under the~orientation reversal. Since the~entire Chern-Simons functional changes sign under parity (orientation reversal), it has the~same effect as changing the~sign of the~``level'' or, equivalently, $q \mapsto q^{-1}$. This is true for any $G=SU(N)$ and, therefore, can be summarised by $(a,q) \mapsto (a^{-1}, q^{-1})$, which is precisely the~statement in \eqref{mirrorHOMFLY}.

Similarly, if we consider Chern-Simons theory with complex gauge group $G_{\mathbb{C}}$ on the~knot complement, $S^3 \setminus K$, then the~parity (orientation reversal) operation induces\footnote{One way to see this is to consider a~simple example, e.g. a~solid torus $\cong S^3 \setminus \text{unknot}$, and realise the~parity transformation as a~sign change of one of the~coordinates along the~boundary.} an~orientation reversal on the~boundary torus $T^2 \cong \partial \left( S^3 \setminus K \right)$. This tells us that out of two $G_{\C}$-valued holonomies $x$ and $y$ along 1-cycles of $T^2$ one should be inverted upon $K \mapsto m(K)$. In particular, when $G_{\mathbb{C}} = SL(2,\mathbb{C})$ this implies the~familiar transformation of the~A-polynomial. Since the~latter is defined only up to overall powers of $x$ and $y$, and enjoys the~Weyl symmetry $A_K (x,y) \sim A_K (x^{-1}, y^{-1})$, it does not matter whether we use $(x,y) \mapsto (x^{-1},y)$ or $(x,y) \mapsto (x,y^{-1})$ as the~effect of the~orientation reversal. The~standard choice is
\begin{equation}
A_{m(K)} (x,y) \; = \; A_{K} (x^{-1},y).
\label{mirrorA}
\end{equation}
Note that the~transformations \eqref{mirrorHOMFLY} and \eqref{mirrorA} are compatible. Indeed, as we reviewed earlier, introducing $a$-dependence and quantising the~classical curve $A_K (x,y)=0$ leads to a~recursion relation $\hat A_K (\hat x, \hat y,a,q) P_* (K;a,q) = 0$ that encodes colour dependence of the~coloured HOMFLY-PT polynomials. According to \eqref{mirrorHOMFLY}, the~recursion relation for the~mirror knot should involve $q^{-1}$ in place of $q$ (accompanied by a~more straightforward transformation $a \mapsto a^{-1}$).
Since $\hat x P_r = q^r P_r$ and $\hat y P_r = P_{r+1}$, it follows that under $q \mapsto q^{-1}$ we have $\hat x \mapsto \hat x^{-1}$ and $\hat y \mapsto \hat y$.
Therefore, the~$(a,q)$-deformed version of \eqref{mirrorA} involves 
$(\hat x, \hat y,a,q) \mapsto (\hat x^{-1}, \hat y,a^{-1},q^{-1})$ under $K \mapsto m(K)$.

Now we are ready to explain why the~relation \eqref{mirrorFF} between the~two functions $F_K (x,a,q)$ and $F_{m(K)} (x,a,q)$ is very nontrivial when $K$ is not amphichiral. Since a~coefficient of each term $a^{n_a} x^{n_x}$ is supposed to be in $\mathbb{Z} [q^{-1}, q]]$, we can not simply take $q \mapsto q^{-1}$. Therefore, from this point of view, $F_K (x,a,q)$ and $F_{m(K)} (x,a,q)$ are very different! For instance, while in this paper we discuss the~form of $F_{5_2} (x,a,q)$, it does not tell us directly what the~corresponding terms in $F_{m(5_2)} (x,a,q)$ should be.

On the~other hand, the~relation between $F_K (x,a,q)$ and $F_{m(K)} (x,a,q)$ at the~perturbative level is very direct and simple.\footnote{Since $\hat y$-coefficients of $\hat A_K (\hat x, \hat y,a,q)$ are rational functions of $\hat x$, $a$, and $q$, in the~quantum A-polynomial one can also simply replace $(\hat x, a, q) \mapsto (\hat x^{-1}, a^{-1}, q^{-1})$.}
As usual, let us write $a = e^{N \hbar}$ and $q = e^{\hbar}$, where $N$ can be treated as a~parameter (not necessarily integer). In these variables, the~mirror transform is $(x,N,\hbar) \mapsto (x^{-1},N,-\hbar)$. Therefore, to all orders in the~perturbative expansion with respect to $a$ and $q$ near $1$, we have
\begin{equation}
F_K (x,a,q) \; = \; \sum_{i,j \ge 0} \sum_{m} c_{i,j,m} \hbar^i N^j x^m
\label{FKpertupper}
\end{equation}
and
\begin{equation}
F_{m(K)} (x,a,q) \; = \; \sum_{i,j \ge 0} \sum_{m} c_{i,j,m} (-\hbar)^i N^j x^{-m}
\label{FKpertlower}
\end{equation}
with the~{\it same} coefficients $c_{i,j,m}$.

The problem, therefore, is to translate a~rather simple relation among the~perturbative coefficients of $F_K (x,a,q)$ and $F_{m(K)} (x,a,q)$ into a~much more sophisticated relation \eqref{mirrorFF}.
Note that this task would be much easier if one had an~{\it a~priori} knowledge about modular properties of $F_K (x,a,q)$ with respect to $q$, i.e. $\tau = \frac{\hbar}{2\pi i}$.
Then, the~modular transformation $\tau \mapsto - \frac{1}{\tau}$ could help relating the~coefficients of the~$q$-series near the~``cusp'' $\tau = i \infty$ to the~perturbative expansion near $q \simeq 1$ or $\tau \simeq 0$. The~tools useful for solving this problem include resurgent analysis \cite{GMP,AP} and Rademacher sums \cite{CCFGH}. 

We next consider the~counterpart of the~above discussion from the~point of view of curve counting. Here the~substitution $(a,q)\to(a^{-1},q^{-1})$ that relates $F_K$ to $F_{m(K)}$ can be derived as follows. Starting from a~knot $K$, we get its mirror knot simply by reversing the~orientation of $S^3$. Reversing the~orientation of $S^3$ reverses the~orientation of the~4-chain of $S^3$ and as the~power of $a$ corresponds to an~intersection number between holomorphic curves and the~4-chain, we see that $a$ should be replaced by $a^{-1}$. Similarly, the~orientation on the~complement Lagrangian $M_K$ changes and since $q$ counts intersections with its 4-chain (corresponding to the~$U(1)$ gauge theory on the~single copy of the~$M_K$-brane), we find that $q$ should be replaced by~$q^{-1}$. 

We next discuss the~coefficients of $a^{n_a}x^{n_x}$ in $F_K$ from the~geometric point of view. We start in the~case of fibered knots. If $K$ is fibered, then $M_K$ can be made disjoint from $S^3$ in $T^\ast S^3$ and moved to the~resolved conifold. In this setting $F_K$ is given by a~count of curves with boundary in $M_K$. Using a~perturbation scheme for counting bare curves as in \cite{ES}, the~contribution of such a~(possibly disconnected) curve $u$ to the~count of generalised curves in $F_K$ has the~form
\begin{equation}\label{eq:curve count}
   w(u)\,(q^{1/2}-q^{-1/2})^{-\chi(u)}\,a^{d(u)/2}q^{l(u)/2}x^{k(u)}, 
\end{equation}
where $w(u)$ is the~rational weight of the~curve as a~point in the~moduli space, $\chi(u)$ is the~Euler characteristic of $u$, $d(u)$ the~homological degree, $l(u)$ a~linking of the~boundary of $u$, and $k(u)$ the~boundary degree. This then means that the~coefficient of $a^{n_a}x^{n_x}$ lies in $\mathbb{Q}[q^{\pm 1/2},(q^{1/2}-q^{-1/2})^{-1}]$ which shows that the~substitution $q\to q^{-1}$ taking us from $F_K$ to $F_{m(K)}$ works well.
Note that the~expression (\ref{eq:curve count}) corresponds to the~{\it fully unreduced} normalisation, natural from the~geometric perspective. The~result in the~reduced normalisation can be obtained by dividing by the~curve count of the~unknot.

If $K$ is not fibered, then -- as explained above -- we must also take into account contributions from curves with additional negative punctures at Reeb chords $\alpha$ connecting fibers corresponding to intersection points in $M_K\cap S^3\subset T^\ast S^3$. The~argument above indicates that the~functions $\alpha=\alpha(a,q)$ transform via the~change of variables $(a,q)\to (a^{-1},q^{-1})$ under change of orientation of $M_K$ and $S^3$. Given this we would have a~similar but somewhat more involved result for the~coefficients as follows. The~coefficients of $a^{n_a}x^{n_x}$ would take values in $\mathbb{Q}[q^{\pm 1/2},(q^{1/2}-q^{-1/2})^{-1}]\otimes\mathbb{Q}[[\alpha(a,q)]]$ with change of variables giving coefficients in $\mathbb{Q}[q^{\pm 1/2},(q^{1/2}-q^{-1/2})^{-1}]\otimes\mathbb{Q}[[\alpha(a^{-1},q^{-1})]]$. Not much is known about the~functions $\alpha(a,q)$ but the~examples from Section \ref{sec:a-deformed F_K} indicate that they could contain rational powers of $a$ and $q$.

We also expect this work to offer a~new territory for studying quantum modular forms and their generalisations. The~notion of quantum modularity, introduced in \cite{Zag}, is about properties of a~function defined only at rational numbers, $\mathbb{Q} \subset \mathbb{R}$, on the~real axis of the~$\tau$-plane. Since in terms of the~variable $q$ these are the~points on the~unit circle, $|q|=1$, naively it seems that $F_K (x,q^N,q)$ and $F_{m(K)} (x,q^N,q)$ discussed earlier have little to do with quantum modularity because they are defined in the~upper half-plane and in the~lower half-plane (or, inside and outside the~unit disk if we use $q$ instead of $\tau$).
However, in all examples (of low rank) that have been studied so far, the~connection to quantum modularity was found by studying (regularised) limits to roots of unity in $q$-variable, i.e. $\tau \to \tau_0 \in \mathbb{Q}$.
Therefore, based on these studies, one might expect that connection to quantum modularity continues to hold more generally.

We can also offer an~intuitive reason why one might expect such connection at roots of unity. Conceptually, $F_K (x,q^N,q)$ and $F_{m(K)} (x,q^N,q)$ are quantum group invariants associated with $\mathcal{U}_q (\mathfrak{sl}_N)$ at generic $|q|<1$. From the~theory of quantum groups and from the~physical realisation of $F_K (x,q^N,q)$, it is clear that (regularised) limits of such functions should be very interesting and contain rich structure if (and, probably, only if) $q \to$ root of 1.
Moreover, at those points ($\tau \in \mathbb{Q}$) the~asymptotic expansions of $F_K (x,q^N,q)$ and $F_{m(K)} (x,q^N,q)$ are expected to be related in a~simple way, cf. \eqref{FKpertupper}--\eqref{FKpertlower}.

Summing up, if the~knot $K$ is not amphichiral,
%To summarise, for a~non-amphichiral knot $K$,
the~two functions $F_K (x,q^N,q)$ and $F_{m(K)} (x,q^{-N},q^{-1})$, naturally defined in the~upper half-plane and in lower half-plane respectively, are in general quite different and related in a~highly nontrivial way, cf. \eqref{mirrorFF}.
Yet, their asymptotic expansions near rational points on the~real axis are related by a~very simple ``analytic continuation'' \eqref{FKpertupper}--\eqref{FKpertlower}.
This peculiar phenomenon, sometimes called ``leaking'', not only provides a~function defined on $\tau \in \mathbb{Q}$, but automatically comes equipped with two analytic continuations of this function to the~upper and lower half-plane. Furthermore, these two functions are expected to have modular properties of characters of logarithmic vertex algebras \cite{CCFGH}. From this perspective, it is perhaps less surprising that limiting values of characters of chiral algebras are related by $SL(2,\mathbb{Z})$ action.

Besides connections to traditional quantum modularity at $a = q^N$, it would be interesting to understand the~properties of $F_K (x,a,q)$ itself. It is quite possible that $F_K (x,a,q)$ is also related to characters of (non-unitary) chiral algebras. We hope that exploring this direction can lead to new types of modularity and strengthen the~connection between enumerative geometry, quantum algebra, and number theory.

\section{Future directions}\label{sec:Future directions}
In this section, we provide a~summary of interesting open problems that emerged during our research:
\begin{itemize}
\item In this paper we noticed a~close relation between $a$-deformed $F_{K}$ invariants and HOMFLY-PT polynomials. For $(2,2p+1)$ torus knots it led to the~closed form expression of $F_K(x,a,q)$ coming from $P_r(K;a,q)$, with $q^r=x$. However, this approach does not work in general: we have seen that for the~figure-eight knot such substitution would lead to an~ill-defined series containing expansion in both $x$ and $x^{-1}$. It would be desirable to solve this problem and understand the~relation between HOMFLY-PT polynomials and $a$-deformed $F_{K}$ invariants in full generality.
\item It would be desirable to prove Conjectures \ref{$a$-deformed $F_K$}-\ref{$a,t$-deformed $F_K$} and define $F_{K}$ in a~proper mathematical way, showing that it is a~topological invariant. Unfortunately, so far this was problematic even in the~simplest $SU(2)$ case. There is, however, a~definition in $SU(2)$ case for positive braid knots \cite{Park2}, and it should not be hard to generalise it to $SU(N)$. It would be interesting to find the~$a$-deformed $F_K$ for positive braid knots using the~same approach, in which the~main step is finding the~$a$-deformed $R$-matrix.
\item We demonstrated how to solve the~recursion to get a~unique solution up to an~overall factor. However, we still do not know how to determine this overall factor, namely the~first coefficient $f_0(a,q)$, especially for non-fibered knots like $5_2$. Clearly we need a~method beyond the~recursion. One possible approach is to combine with the~expected asymptotic series \eqref{eq:asymptotics}. Another approach would be to use the~$a$-deformed $R$-matrix, if it exists. 
\item The~close relationship between $a$-deformed $F_{K}$ invariants and HOMFLY-PT polynomials suggests that the~knots-quivers correspondence \cite{KRSS1, KRSS2} can be generalised to the~case of knot complements. The first results presented in \cite{Kuch} seem to confirm this hypothesis.
\item Since knot complements can be glued to give a~closed 3-manifold, Dehn surgery of $F_{K}$ invariants leads to $\widehat{Z}$ invariants \cite{GM, Park1}. It would be interesting to find a~large $N$ limit and $t$-deformation of this relation. However, for this we need $F_K$ that takes care of all possible Young diagrams, not just symmetric ones,  i.e. allows generic values of all the~variables $x_i$ in \eqref{FKZhat}. In the~long term one may hope that such developments will be helpful in the~categorification of the~Witten-Reshetikhin-Turaev invariants.

\item As a~step toward exploring the~relation between $t$-deformation and categorification of $F_K$ invariants, it would be interesting to study a~similar relation between $t$-deformation and categorification of ADO$_p (x)$ polynomials that arise as limits of $F_K (x,q)$ at roots of unity. We hope that our computations of $t$-deformed ADO polynomials in Section~\ref{sec:t-deformation} will be useful for carrying out this analysis.

\item It would be interesting to identify chiral algebras that may have $F_K (x,a,q)$ as their characters. Addressing this is closely related to understanding the~modular properties of $F_K (x,a,q)$ as well as its relation to $F_{m(K)} (x,a,q)$. All these questions, that we leave to future work, are intimately interrelated.

\item Curiously, both the~enumerative perspective discussed here and potential interpretation of $F_K (x,a,q)$ as characters of logarithmic VOAs suggest that $F_K (x,a,q)$ should satisfy $q$-difference equation with respect to variable $a$,  i.e. $q$-difference equations where $a$ plays the~role similar to that of $x$ and the~``shift operator'' acts as $a^n \mapsto q^n a^n$. 
\end{itemize}

\section*{Acknowledgments}
We would like to thank Sibasish Banerjee, Miranda Cheng, Luis Diogo, Boris Feigin, Francesca Ferrari, Sarah Harrison, Jakub Jankowski, Pietro Longhi, Ciprian Manolescu, Marko Sto$\check{\text{s}}$i$\acute{\text{c}}$, Cumrun Vafa, and Don Zagier for insightful discussions and comments on the~draft. The~work of T.E. is supported by the~Knut and Alice Wallenberg Foundation and the~Swedish Research Council. The~work of S.G. is supported by the~U.S. Department of Energy, Office of Science, Office of High Energy Physics, under Award No. DE-SC0011632, and by the~National Science Foundation under Grant No. NSF DMS 1664240. The~work of P.K. is supported by the~Polish Ministry of Science and Higher Education through its programme Mobility Plus (decision no. 1667/MOB/V/2017/0). The~research of S.P. is supported by Kwanjeong Educational Foundation. The~work of P.S. is supported by the~TEAM programme of the~Foundation for Polish Science co-financed by the~European Union under the~European Regional Development Fund (POIR.04.04.00-00-5C55/17-00).

\newpage 

\bibliography{HOMFLYPTFK}
\bibliographystyle{abstract}

\end{document}